\documentclass{article}

\usepackage{arxiv}
\usepackage{amssymb}
\usepackage{amsmath}
\usepackage{empheq}
\usepackage{amscd}
\usepackage{graphicx}
\usepackage{graphics}
\usepackage[noadjust]{cite}
\usepackage{amsthm}
\usepackage{caption}
\usepackage{multirow}
\usepackage{array}
\usepackage{rotating}

\usepackage{indentfirst}
\usepackage{tikz}
\usepackage{calc}
\usepackage{epsfig}
\usepackage{natbib}
\usepackage[makeroom]{cancel}
\usepackage{multicol}
\usepackage{csquotes}
\usepackage{algorithm}
\usepackage{algorithmic}
\usepackage{enumitem}
\usepackage{hyperref}
\usepackage{url}
\hypersetup{colorlinks,linkcolor={blue},citecolor={blue},urlcolor={blue}}

\usepackage{timet}
\usepackage{graphicx}
\usepackage{multirow}
\usepackage{multicol}
\usepackage{lipsum}
\usepackage{timet}
\usepackage{epsfig}
\usepackage{amsmath}
\usepackage{amsfonts}
\usepackage{amssymb}
\usepackage{color}
\numberwithin{equation}{section}
\usepackage{caption}
\usepackage{float}
\usepackage{subcaption}
\usepackage{graphics}
\usepackage{xcolor,graphicx}
\usepackage{csquotes}
\usepackage{mathtools}
\usepackage{optidef}

\newcommand{\vertiii}[1]{{\left\vert\kern-0.25ex\left\vert\kern-0.25ex\left\vert #1 
    \right\vert\kern-0.25ex\right\vert\kern-0.25ex\right\vert}}
    
\usepackage{array,multirow,makecell}
\newcolumntype{R}[1]{>{\raggedleft\arraybackslash }b{#1}}
\newcolumntype{L}[1]{>{\raggedright\arraybackslash }b{#1}}
\newcolumntype{C}[1]{>{\centering\arraybackslash }b{#1}}

\newcommand{\ket}{\right\rangle}       
\newcommand{\bra}{\left\langle}

\def\tsc#1{\csdef{#1}{\textsc{\lowercase{#1}}\xspace}}
\tsc{WGM}
\tsc{QE}
\tsc{EP}
\tsc{PMS}
\tsc{BEC}
\tsc{DE}

\begin{document}
\title{Full waveform inversion beyond the Born approximation:
A tutorial review}

\author{
\href{http://orcid.org/0000-0002-4981-4967}{\includegraphics[scale=0.06]{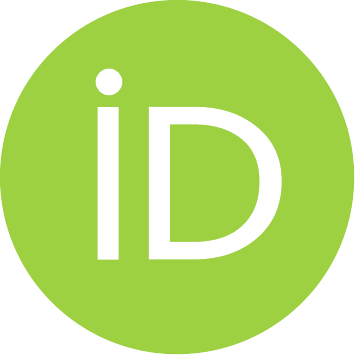}\hspace{1mm}St\'ephane Operto} \\ 
  University Cote d'Azur - CNRS - IRD - OCA, Geoazur, Valbonne, France. 
  \texttt{operto@geoazur.unice.fr} 
\And
 \href{https://orcid.org/0000-0002-9879-2944}{\includegraphics[scale=0.06]{orcid.pdf}\hspace{1mm}Ali Gholami} \\
  Institute of Geophysics, University of Tehran, Tehran, Iran.
  \texttt{agholami@ut.ac.ir} \\ 
\And
\href{http://orcid.org/0000-0003-1805-1132}{\includegraphics[scale=0.06]{orcid.pdf}\hspace{1mm}Hossein S. Aghamiry} \\
  University Cote d'Azur - CNRS - IRD - OCA, Geoazur, Valbonne, France. 
  \texttt{aghamiry@geoazur.unice.fr}
\And 
 \href{https://orcid.org/0000-0002-9879-2944}{\includegraphics[scale=0.06]{orcid.pdf}\hspace{1mm}Gaoshan Guo} \\
  University Cote d'Azur - CNRS - IRD - OCA, Geoazur, Valbonne, France.
  \texttt{guo@geoazur.unice.fr} \\ 
\And
 \href{https://orcid.org/0000-0002-9879-2944}{\includegraphics[scale=0.06]{orcid.pdf}\hspace{1mm}Frichnel Mamfoumbi} \\
  University Cote d'Azur - CNRS - IRD - OCA, Geoazur, Valbonne, France.
  \texttt{mamfoumbi@geoazur.unice.fr} \\ 
  \And
 \href{https://orcid.org/0000-0002-9879-2944}{\includegraphics[scale=0.06]{orcid.pdf}\hspace{1mm}Stephen Beller} \\
  University Cote d'Azur - CNRS - IRD - OCA, Geoazur, Valbonne, France.
  \texttt{beller@geoazur.unice.fr} \\   
 }

\renewcommand{\shorttitle}{FWI beyond Born approximation, Operto et al.}

\maketitle

\begin{abstract}
Full Waveform Inversion can be made immune to cycle skipping by matching the recorded data arbitrarily well from inaccurate subsurface models. To achieve this goal, the simulated wavefields can be computed in an extended search space as the solution of an overdetermined problem aiming at jointly satisfying the wave equation and fitting the data in a least-squares sense. Simply put, the wavefields are computed by solving the wave equation in the inaccurate background model with a feedback term to the data added to the physical source in the right-hand side. Then, the subsurface parameters are updated by canceling out these additional source terms, sometime called unwisely wave-equation errors, to push the background model towards the true model in the left-hand side wave-equation operator. Although many studies were devoted to these approaches with promising numerical results, their governing physical principles and their relationships with classical FWI don't seem to be understood well yet. The goal of this tutorial is to review these principles in the framework of inverse scattering theory whose governing forward equation is the Lippmann-Schwinger equation. From this equation, we show how the data-assimilated wavefields embed an approximation of the scattered field generated by the sought model perturbation and how they modify the sensitivity kernel of classical FWI beyond the Born approximation. We also clarify how the approximation with which these wavefields approximate the unknown true wavefields is accounted for in the adjoint source of the parameter estimation problem. The theory is finally illustrated with numerical examples. Understanding the physical principles governing these methods is a necessary prerequisite to assess their potential and limits and design relevant heuristics to manage the latter.
\end{abstract}

\graphicspath{{./figs/}}
\section{Introduction}
\begin{figure}[htb!] \label{fig_algo}
\centering
\begin{center}
\includegraphics[width=14cm,clip=true,trim=0cm 0cm 0.cm 0cm]{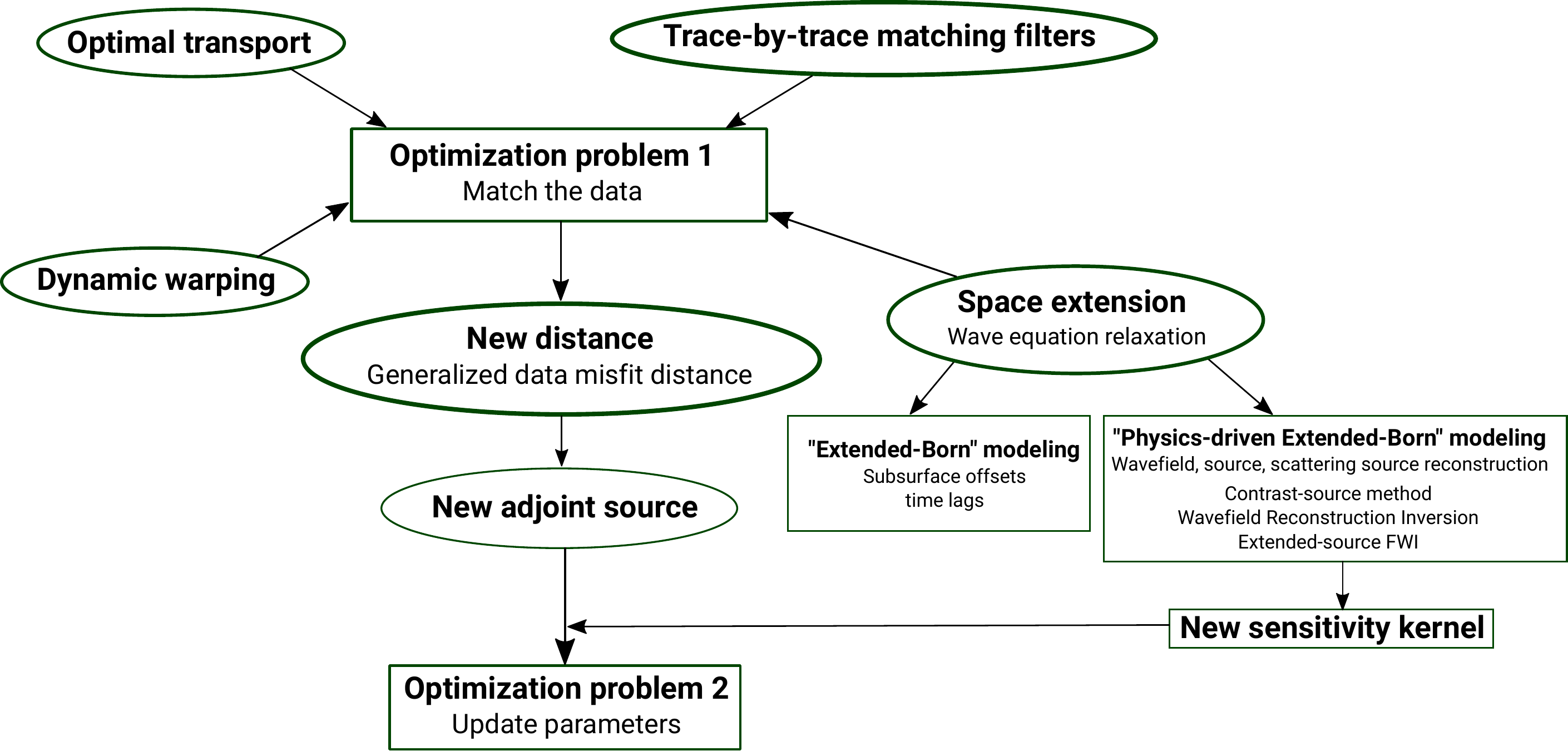} 
\end{center}
\caption{Sketch of new formulations of FWI aiming at minimizing cycle skipping. A first optimization subproblem is solved to build a more convex distance than the least-squares data misfit function by matching the observables well enough. This can be performed by optimal transport, dynamic warping, matching filters, and approximated wave equation relaxation. In the later case, the extended search space can be implemented in the model domain (Extended-Born modeling methods) or in the source domain (contrast-source, wavefield reconstruction, extended-source methods). These approaches modify the adjoint source of the parameter-estimation subproblem. Some approaches also modify the sensitivity kernel of classical FWI.} 
\end{figure}

\noindent Full Waveform Inversion (FWI) \citep{Tarantola_1984_ISR,Gauthier_1986_TDN,Mora_1987_NTD,Pratt_1998_GNF} has become the baseline seismic imaging method to build high-resolution multi-parameter subsurface models from full-azimuth long-offset data since a decade \citep{Sirgue_2010_FWI}. 
Despite its popularity, FWI remains a highly nonlinear problem due to the oscillatory nature of seismic signals, which makes the classical least-squares data-fitting misfit function highly multimodal. In this context, a central issue of FWI is cycle skipping, which traps FWI in spurious minima as soon as the simulated data don't predict the recorded counterpart with an error lower than half a period \citep{Virieux_2009_OFW}. Remembering that FWI is an inverse scattering problem solved with gradient-based methods, this stringent condition results from the weak-scattering Born approximation with which the nonlinear data fitting problem is linearized around a background model to recast it as a sequence of surrogate linear problems.
On the acquisition side, sources with low-frequency content ($\ge$ 1.5~Hz) have been designed to mitigate cycle skipping \citep[e.g.][]{Baeten_2013_ULF,Brenders_2022_WEL}. On the imaging side, new FWI formulations match the data arbitrarily well from inaccurate velocity models, hence preventing cycle skipping. To achieve this goal, a first optimization subproblem is solved to build a more convex distance in place of the usual least-squares data misfit distance before minimizing this new distance for parameter estimation (Figure~\ref{fig_algo}). Several approaches can be viewed to match the data with inaccurate models such as optimal transport \citep[e.g.,][]{Engquist_2016_WAS,Yang_2017_AOT,Metivier_2018_OTM}, trace-by-trace matching filters \citep[e.g.,][]{Warner_2016_AWI}, dynamic warping based approaches \citep[e.g.,][]{Ma_2013_WRT} and error in constitutive equation approaches \citep[e.g.,][]{Banerjee_2013_LSP}. 
The later extend the search space of classical FWI by adding some degrees of freedom in the forward-problem equation to match the data with the desired accuracy. 
These approaches include the so-called Extended-Born modeling approaches \citep[e.g][]{Symes_2008_MVA,Biondi_2014_SIF,Barnier_2022_FWI}, the contrast-source method \citep[e.g.,][]{vandenBerg_1997_CSI,Abubakar_2009_FDC}, the wavefield reconstruction inversion method \citep[e.g.,][]{VanLeeuwen_2013_MLM,vanLeeuwen_2016_PMP} and the extended-source FWI \citep[e.g.,][]{Wang_2016_FIR,Huang_2018_SEW}. Extended-Born modeling approaches extends the search space in the model domain with subsurface offsets or time lags, while the last three methods, which mainly differ in the parametrization of the optimization variables, extend the search space in the source domain. \\
%
%

\noindent All these approaches modify the source of the adjoint-state equation in the descent direction of the model updating and hence can be recast as a generalized FWI \citep{Tarantola_1982_GNI}, where a data-domain covariance matrix in the least-squares data-fitting misfit function transforms the data residuals such that the weighted misfit function is more convex. Some of them also modify the sensitivity kernel of FWI when the covariance matrix depends on the model parameters as we will show in this study. \\
\noindent This study focuses on the error in constitutive equation approaches where the search space extension is implemented in the source domain, which will be called extended-space (ES) FWI. After the seminal paper of \citet{VanLeeuwen_2013_MLM}, ES-FWI have been investigated in several studies (Table~\ref{tab_ref}) with however specific angles of approach of mathematical, algorithmic or applied natures but none of them provided a comprehensive overview of the physical principles governing these methods and their underlying approximations. The goal of this tutorial is to fill this gap by gathering the lessons learnt from these studies into a self-contained review of these principles.
By clarifying these physical principles, we will clearly draw the similarities and differences between classical FWI and ES-FWI. 
That is, we will show that ES-FWI relies on the same diffraction-based physical principles as those governing classical FWI, while mitigating the detrimental effects of the Born approximation with more accurate approximation of the wavefields.
Many inverse scattering problems revolve around the Lippmann-Schwinger equation giving the scattered field by a model perturbation. Accordingly, this equation will be the cornerstone of this tutorial. Let's add that this preserved connection between ES-FWI and the physics of diffraction somehow contrasts with other approaches that are more inspired by signal or image processing like matching filters and dynamic warping based approaches or by optimization theory like optimal transport. Finally, we would like to summarize the governing idea of ES-FWI by borrowing two sentences from the preface of \citet{Chavent_1996_IPW} introducing the approximate inverse scattering method of \citet{Fiddy_1996_RSS}. In a similar manner to ES-FWI, this method aims at improving the ability of inverse scattering theory based on the Born approximation to reconstruct strong scattering objects from the Lippmann-Schwinger equation: \textit{The method is better than Born one but it keeps the physical features that appear in this approximate method. It goes also as close as possible to exact methods without losing physical features}.

\noindent In the theory section, we first review the bilinearity of the wave equation to remind that ES-FWI aims at reconstructing wavefields that best approximate the true ones. Second, we review the Lippmann-Schwinger equation and the Born approximation before reminding their role in classical FWI. Then, we explain how improved wavefields can be reconstructed from the Lippmann-Schwinger equation and the recorded scattered data before explaining how they modify the sensitivity kernel of FWI. Finally, we explain how the approximation with which these wavefields are reconstructed is accounted for in the adjoint source of ES-FWI. We briefly review some important numerical aspects of ES-FWI to design workable algorithms before illustrating the theory with two numerical examples.

\begin{center}
\begin{table}[ht!]
\centering
\begin{tabular}{|L{4.5cm}|L{9.5cm}|}
\hline
\small{\citet{Chavent_1996_IPW}} & \small{Lecture notes on Inverse problems of wave propagation and diffraction.}  \\  \hline  
\small{\citet{vandenBerg_1997_CSI}} & \small{Contrast source inversion method for strongly scattering objects.}  \\  \hline
\small{\citet{Abubakar_2008_FCS,Abubakar_2009_FDC,Abubakar_2011_TSF}} 	& \small{Application of contrast-source method to seismic waveform inversion.}  	\\ \hline                       
\small{\citet{VanLeeuwen_2013_MLM}}  &  \small{Wavefield Reconstruction Inversion (WRI) with alternating directions.} \\ \hline
\small{\citet{vanLeeuwen_2014_NTF}}  &  \small{Tentative analogy between source extension and adjoint wavefield.} \\ \hline
\small{\citet{vanLeeuwen_2016_PMP}}   &  \small{Reduced form of WRI (variable projection) and mathematical aspects.} \\ \hline
\small{\citet{Wang_2016_FIR,Wang_2017_RFI}} & \small{Time-domain WRI with approximated extended source.} \\  \hline
\small{\citet{Fu_2017_DPM}}  & \small{Discrepancy-based penalty method for adaptive penalty parameter.} \\  \hline
\small{\citet{Huang_2018_SEW}}   & \small{Source-signature independent extended-source FWI.} \\ \hline
\small{\citet{Huang_2018_VSE}}   & \small{Volume extended-source FWI (first temporal sample of extended source).} \\ \hline
\small{\citet{Fang_2018_SEF}} & \small{Source signature estimation in WRI.} \\ \hline
\small{\citet{Aghamiry_2019_IWR}} & \small{Frequency-domain WRI with Augmented Lagrangian (AL) (IR-WRI).} \\ \hline
\small{\citet{Aghamiry_2019_AMW}} & \small{Extension of IR-WRI to VTI media.} \\ \hline
\small{\citet{Aghamiry_2019_AMW,Aghamiry_2019_CRO}} & \small{TV $\&$ compound regularizations, and bound constraints in IR-WRI.} \\ \hline
\small{\citet{Peters_2019_ANS}} & \small{Frequency-domain wavefield reconstruction in 3D large-scale models.} \\ \hline
\small{\citet{vanLeeuwen_2019_ANO}}   & \small{Recognize WRI as a generalized form of FWI.} \\ \hline
\small{\citet{Aghamiry_2020_RWI}} & \small{IR-WRI with phase retrieval.} \\ \hline
\small{\citet{Aghamiry_2019_VWR,Aghamiry_2020_CIT}} & \small{IR-WRI for attenuation imaging.} \\ \hline
\small{\citet{Symes_2020_WRI}} & \small{Mathematical analysis of WRI convexity.} \\ \hline
\small{\citet{Lee_2020_SFW}} & \small{Weighted scattering source minimization in WRI.} \\ \hline
\small{\citet{Aghamiry_2021_EES}} & \small{Source signature estimation in IR-WRI.} \\ \hline
\small{\citet{Aghamiry_2021_OEF}} & \small{On the analogy between the contrast source method and WRI.} \\ \hline
\small{\citet{Rizzuti_2021_DFW}} & \small{Dual formulation of WRI in the data domain.} \\ \hline
\small{\citet{Gholami_2022_EFW}} & \small{Efficient data-domain algorithm for time-domain IR-WRI.} \\ \hline
\small{\citet{Aghazade_2022_AAA}} & \small{IR-WRI with Anderson acceleration.} \\ \hline
\small{\citet{Aghazade_2021_RSS}} & \small{IR-WRI with randomized source sketching.} \\ \hline
\small{\citet{Hajjaj_2022_wavefield}} & \small{Wavefield reconstruction in the framework of Marchenko theory.} \\ \hline
\small{\citet{Lin_2022_TWR}} & \small{Data-domain Hessian with point-spread function.} \\ \hline
\small{\citet{Guo_2022_PID}} & \small{Data-domain Hessian with matching filter $\&$ truncated iterative method.} \\ \hline
\end{tabular}
\caption{Main references about ES-FWI since the first publications on the contrasted source method.} 
\label{tab_ref}
\end{table}
\end{center}


\section{Theory}

\subsection{On the bilinearity of the wave equation}
\noindent To introduce the governing idea of ES-FWI, we may begin with reminding the bilinearity of the wave equation.  In this study, we will restrict ourselves to the Helmholtz equation written in matrix form as:
\begin{equation}
\bold{A}(\bold{m}) \bold{u} = \bold{b},
\label{eqA1}
\end{equation}
\noindent where $\bold{A}(\bold{m})=\omega^2 \text{diag}(\bold{m}) + \nabla^2$ is the so-called impedance matrix, $\omega$ is the angular frequency, $\text{diag}(\bullet)$ is a diagonal matrix formed with the coefficients of vector $\bullet$, model parameters $\bold{m}$ are squared slowness, $\nabla^2$ is the Laplace operator, $\bold{u}$ is the monochromatic wavefield and $\bold{b}$ is the source. Moreover,  $\bullet_i$ denotes the $i^{th}$ entry of the vector $\bullet$ through the rest of the tutorial.

\begin{figure}[htb!]
\centering
\begin{center}
\includegraphics[width=14cm,clip=true,trim=0cm 0cm 0.cm 0cm]{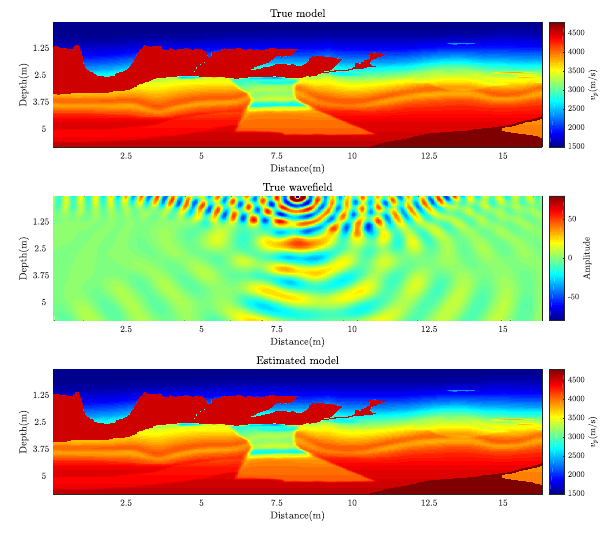} 
\caption{Bilinearity of the wave equation. (a) 2004 BP salt model. (b) Three-Hertz monochromatic wavefield triggered by a source in the middle of the model. (c) Reconstructed model from equation~\ref{eqbili} \citep{Aghamiry_2019_IWR}.}
\label{fig_bilinearity}
\end{center}
\end{figure}

\noindent Equation \ref{eqA1} is linear in $\bold{u}$ for known $\bold{m}$. Permuting $\bold{m}$ and  $\bold{u}$ in the mass term, $\text{diag}(\bold{m}) \bold{u}=\text{diag}(\bold{u}) \bold{m}$, recasts equation~\ref{eqA1} as
\begin{equation}
\bold{L}(\bold{u}) \bold{m} = \bold{y}(\bold{u}),
\label{eqA2}
\end{equation}
\noindent where $\bold{L}(\bold{u})=\omega^2 \text{diag}(\bold{u})$ and $\bold{y}(\bold{u})=\bold{b} - \nabla^2 \bold{u}$.
\noindent Equation~\ref{eqA2} is affine in $\bold{m}$ for known $\bold{u}$ and $\bold{m}$ can be inferred from $\bold{u}$ by a pointwise division since $\bold{L}(\bold{u})$ is a full-rank diagonal matrix:
\begin{equation}
\bold{m} = \frac{\bold{y}(\bold{u})}{\bold{L}(\bold{u})}.
\label{eqbili}
\end{equation}
\noindent This implies that, if we could record a monochromatic wavefield triggered by a monochromatic source everywhere in the subsurface, then we could reconstruct $\bold{m}$ exactly (Figure~\ref{fig_bilinearity}).
This condition is indeed never satisfied since data are typically recorded near the surface, which makes FWI an underdetermined problem. Accordingly, fitting the data at receivers with an error smaller than half the period is a necessary condition to avoid cycle skipping but it is not sufficient to guarantee convergence toward the global minimizer since the wavefield should be matched not only at receivers but also everywhere in the targeted domain. \\


\noindent Accordingly, the objective of ES-FWI is to estimate wavefields that best  match the true wavefields by solving a first optimization problem before updating the subsurface parameters from these improved wavefields. Since the true wavefields cannot be estimated in one iteration due to the underdetermined nature of the problem, this two-step cycle is typically iterated until convergence.\\

\noindent This objective raises the underlying questions: 

\begin{enumerate}

\item How to estimate the best-fitting wavefields from parsimonious data and inaccurate $\bold{m}$?

\item How accurately do they match the true wavefields? 

\item How do they modify the sensitivity kernel of classical FWI? 

\item How does the (limited) accuracy with which they match the true wavefields accounted for in the so-called adjoint source of the FWI gradient? 

\end{enumerate}

\noindent Answering these questions will draw the similarities and differences between FWI and ES-FWI.

\subsection{FWI as a constrained optimization problem: full space versus reduced space methods}

\noindent Before answering these questions, we review how the search-space extension is implemented in ES-FWI from the optimization viewpoint. Broadly defined, FWI is a partial differential equation constrained optimization problem
\begin{equation}
\min_{\bold{u}_s,\bold{m}} \sum_{s=1}^{n_s} \| \bold{P}  \bold{u}_s - \bold{d}_s^* \|_2^2 ~~~ \text{subject to} ~~~ \bold{A}(\bold{m})\bold{u}_s-\bold{b}_s = \bold{0}, s=1,...,n_s, 
\label{eqconstrain}
\end{equation}
\noindent where $n_s$ denotes the number of sources, $\bold{d}_s^*$ the recorded data for source $s$, and the optimization variables gather the wavefields $\bold{u}_s$ and the subsurface parameters $\bold{m}$. We assume a stationary-recording acquisition ($\bold{P}$ doesn't depend on $s$) for sake of compact notation but the method can be applied to other acquisition geometries.

\noindent In the full space approach, this constrained optimization problem can be solved iteratively with the Lagrange multiplier method
\begin{equation}
\min_{\bold{u}_s,\bold{m}} \max_{\bold{v}_s} \sum_{s=1}^{n_s} \| \bold{P}  \bold{u}_s - \bold{d}_s^* \|_2^2 + \sum_{s=1}^{n_s} \bra \bold{v}_s, \bold{A}(\bold{m})\bold{u}_s-\bold{b}_s\ket_{\mathcal{U}}, \label{al}
\end{equation}
\noindent where $\bold{v}_s$ denotes the Lagrange multipliers, $\bra \cdot ,\cdot  \ket$ the inner product, and $\mathcal{U}$ the state space. The solution of this multivariate problem is found at the saddle point of the above Lagrangian function.
When this constrained optimization problem is solved with Newton algorithms, $\bold{u}_s$, $\bold{m}$ and $\bold{v}_s$ are updated jointly at each iteration by solving a normal-equation system, the so-called Karush-Kuhn-Tucker (KKT) system,  whose unknowns gather the multivariate descent direction, the right-hand side is the multivariate gradient and the left operator is the multivariate Hessian \citep[e.g.,][]{Epanomeritakis_2008_NCG}. This linear system is prohibitively expensive to solve due to its size.

\noindent In classical FWI, the wave-equation constraint is strictly satisfied at each iteration, namely the full space spanned by the three classes of variables is projected onto the parameter space. Accordingly,  the wavefields $\bold{u}_s$ are eliminated from the optimization variables in equation~\ref{eqconstrain} by enforcing their closed-form expression as a function of $\bold{m}$ in the data-misfit function following a variable-projection approach \citep{Golub_2013_VPM}. This leads to the monovariate data-fitting problem
\begin{equation}
\min_{\bold{m}} \phi^{FWI}(\bold{m})=\sum_{s=1}^{n_s}  \| \bold{S}(\bold{m})  \bold{b}_s - \bold{d}_s^* \|_2^2,
\label{eqfwimf}
\end{equation}
\noindent where $\bold{S}(\bold{m}) = \bold{P} \bold{A}^{-1}(\bold{m})$ is the forward modeling operator. As already mentioned, this data-fitting misfit function is highly multimodal due to its sensitivity to kinematic errors. \\

\noindent In ES-FWI, the search space of classical FWI is extended by re-introducing the wavefields as optimization variables \citep{VanLeeuwen_2013_MLM,vanLeeuwen_2016_PMP}. This is implemented by processing the wave equation as a soft constraint with penalty methods, equation~\ref{penalty}, or augmented Lagrangian methods, equation~\ref{al}:
\begin{eqnarray}
\min_{\bold{u}_s,\bold{m}} \phi^P(\bold{m},\bold{u}_s)  =  \sum_{s=1}^{n_s} \| \bold{P}  \bold{u}_s - \bold{d}_s^* \|_2^2  & + & \mu \sum_{s=1}^{n_s} \| \bold{A}(\bold{m})\bold{u}_s-\bold{b}_s \|_2^2, \label{penalty}\\
\min_{\bold{u}_s,\bold{m}} \max_{\bold{v}_s} \phi^{AL}(\bold{m},\bold{u}_s,\bold{v}_s)  = \ \sum_{s=1}^{n_s} \| \bold{P}  \bold{u}_s - \bold{d}_s^* \|_2^2 & + &\mu \sum_{s=1}^{n_s} \| \bold{A}(\bold{m})\bold{u}_s-\bold{b}_s \|_2^2 \nonumber \\
& + & \sum_{s=1}^{n_s} \bra \bold{v}_s, \bold{A}(\bold{m})\bold{u}_s-\bold{b}_s\ket_{\mathcal{U}}, \label{al}
\end{eqnarray}
\noindent where $\mu \in \mathbb{R}_+$ is a penalty parameter.
Compared to the penalty function, the Augmented Lagrangian function incorporates a second leverage (the Lagrangian term) that helps to satisfy the constraint accurately at the convergence point even with a fixed $\mu$ through the defect correction action of the Lagrange multipliers $\bold{v}_s$. We will review the theoretical aspects of ES-FWI with the penalty formulation for sake of simplicity and because it is enough to understand the physical principles governing ES-FWI beyond the Born approximation. However, the numerical experiments presented at the end of this study will be performed with the Augmented Lagrangian implementation to illustrate the real potential of ES-FWI. The reader is referred to \citet{Aghamiry_2019_IWR} and \citet{Gholami_2022_EFW} for more details about the frequency-domain and time-domain implementations of augmented-Lagrangian based ES-FWI. \\

\noindent From the algorithmic viewpoint, $\bold{u}_s$ and $\bold{m}$ are not updated jointly because this would be prohibitively computationally expensive. Instead, they are updated in an alternating mode \citep{VanLeeuwen_2013_MLM} or by variable projection of the wavefields in the parameter-estimation subproblem like in classical FWI \citep{vanLeeuwen_2016_PMP}. In the former case, the $\bold{m}$-subproblem is linear by virtue of the bilinearity of the wave equation. 
The relaxation of the wave equation generated by the penalty function implies that the wavefields in the extended search space, referred to as $\bold{u}^e_s$ where the subscript $^e$ stands for \textit{extended}, satisfy the wave equation with some errors $\delta \bold{b}^e_s$:
\begin{equation}
\bold{A}(\bold{m})\bold{u}^e_s = \bold{b}_s + \delta \bold{b}^e_s.
\end{equation}
\noindent These errors have however a clear physical meaning in the framework of diffraction theory as we will explain later.
Moreover, the penalty function, equation~\ref{penalty}, implies that each wavefield $\bold{u}^e_s$ is the solution of an overdetermined linear system gathering the observation equation $\bold{P}  \bold{u}_s - \bold{d}_s^*$ and the wave equation \citep[][ their equation 6]{VanLeeuwen_2013_MLM}. Accordingly, $\bold{u}^e_s$ are reconstructed with a feedback to the data $\bold{d}_s^*$ and, hence $\delta \bold{b}^e_s$ should involve $\bold{d}_s^*$ in one way or another. Accordingly, we call  $\bold{u}^e_s$ data-assimilated wavefields.

\noindent The estimation of the source extensions $\delta \bold{b}^e_s$, their relationship with the recorded data $\bold{d}_s^*$ and their key role in the parameter estimation are reviewed in the sequel of this tutorial.



%
%

\subsection{The Lippmann-Schwinger equation and the Born approximation}
\noindent Understanding how data-assimilated wavefields $\bold{u}^e_s$ approximate the true wavefields $\bold{u}_s^*$ and the role of the source extensions $\delta \bold{b}_s^e$ in these wavefield reconstructions requires to review the Lippmann-Schwinger equation \citep[e.g.,][]{Prunty_2020_ALI}.

\begin{figure}[htb!]
\centering
\begin{center}
\includegraphics[width=12cm,clip=true,trim=0cm 9cm 0.cm 0cm]{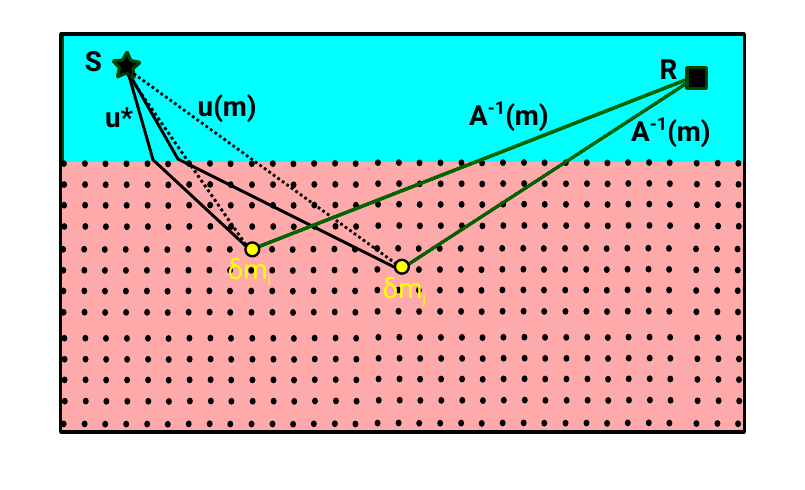} 
\caption{Sketch of the volume integral Lippmann-Schwinger equation. The ground-truth model $\bold{m}^*$ contains two homogeneous layers while the background  medium $\bold{m}$ is homogeneous with the velocity of the top layer. Therefore, the domain of integration is the bottom layer as sketched by the fine grid of scatterers (black circles). The volume source is formed by sampling the true wavefields $\bold{u}^*$ with the scatterers $\delta \bold{m}_i$. The scattered field is then propagated from this source by the Green functions in $\bold{m}$, namely $\bold{A}^{-1}(\bold{m})$ (green lines). In the Born approximation, the incident field to the scatterers is the background wavefield $\bold{u}(\bold{m})$ (dot black lines) instead of $\bold{u}^*$ (solid black lines) leading to a poor approximation of the scattered field (both from the kinematic and dynamic viewpoints) when strong scattering is generated by the perturbation model. The labels $S$ and $R$ denote the source and the receiver, respectively.}
\label{fig_lsequation}
\end{center}
\end{figure}

\noindent Let's denote  the ground-truth model and the background model by $\bold{m}^*$ and $\bold{m}$, respectively. The true wavefield $\bold{u}^*$ and the background wavefield $\bold{u}$ satisfy the wave equation 
\begin{eqnarray}
\bold{A}(\bold{m}^*) \bold{u}^* & = &\bold{b}, \label{eqwaveeqtrue} \\
\bold{A}(\bold{m}) \bold{u} & = & \bold{b}. \label{eqwaveeqback}
\end{eqnarray}
\noindent Moreover, $\bold{u}^*$ can be written as the sum of $\bold{u}$ and the true scattered wavefield $\delta \bold{u}^*$ by the true perturbation model or scattering object $\delta \bold{m}^* = \bold{m}^* - \bold{m}$:
\begin{equation}
\bold{u}^* = \bold{u} + \delta \bold{u}^*.
\label{equ*vu}
\end{equation}
\noindent Equating equations \ref{eqwaveeqtrue} and \ref{eqwaveeqback} and substituting $\bold{u}$ by $\bold{u}^*-\delta \bold{u}^*$, equation~\ref{equ*vu}, gives the Lippmann-Schwinger integral equation satisfied by $\delta \bold{u}^*$
\begin{eqnarray}
\bold{A}(\bold{m}) \delta \bold{u}^* & = & - \left(\bold{A}(\bold{m}^*) - \bold{A}(\bold{m}) \right) \bold{u}^*  \nonumber \\
                                     & = & - \omega^2 \text{diag}(\delta \bold{m}^*) \bold{u}^* \nonumber \\
                                     & = & -\omega^2 \sum_i w_i \boldsymbol{\delta}(\bold{x}-\bold{x}_i) \nonumber \\
                                     & = & \delta \bold{b}^*,
\label{eqdu*}
\end{eqnarray}
\noindent where $w_i = \delta \bold{m}_i^* \bold{u}_i^*$ and $\delta \bold{b}^*$ is the true scattering source.
\noindent This equation shows that the true scattered wavefield  $\delta \bold{u}^*$ is the weighted superposition of the impulse response of the background $\bold{m}$ over the region containing $\delta \bold{m}^*$ where the weights $w_i$ are the point wise product between $\delta \bold{m}^*$ and $\bold{u}^*$ incident to the scatterers \citep{Prunty_2020_ALI}. The ingredients involved in this volume integral equation (namely, $\bold{u}^*$, $\delta \bold{m}^*$ and the Green functions $\bold{A}^{-1}(\bold{m})$)  are sketched in Figure~\ref{fig_lsequation} for a two-layer medium.

%
\noindent Assuming for now that $\delta \bold{b}^*$ is known, we can compute the true wavefield $\bold{u}^*$ by solving the wave equation in the background $\bold{m}$ with an extended source $\bold{b} + \delta \bold{b}^*$:
\begin{equation}
\bold{A}(\bold{m}) \bold{u}^* = \bold{b} + \delta \bold{b}^*.
\label{equ*1}
\end{equation}
\noindent In inverse scattering problems, the scattered wavefields are classically computed with the Born approximation by replacing the true wavefield $\bold{u}^*$ by the background wavefield $\bold{u}(\bold{m})$ in the weights $w_i$ of the scattering source $\delta \bold{b}$  (Figure~\ref{fig_lsequation}). This approximation is only valid for weak scattering. Therefore, ES-FWI aims at extending the linear regime of FWI beyond the Born approximation to avoid cycle skipping by estimating an improved approximation of the scattering source $\delta \bold{b}^*$.

\subsection{The Lippmann-Schwinger equation and classical FWI}
\begin{figure}[htb!]
\centering
\begin{center}
\includegraphics[width=9cm,clip=true,trim=0cm 0cm 0.cm 0cm]{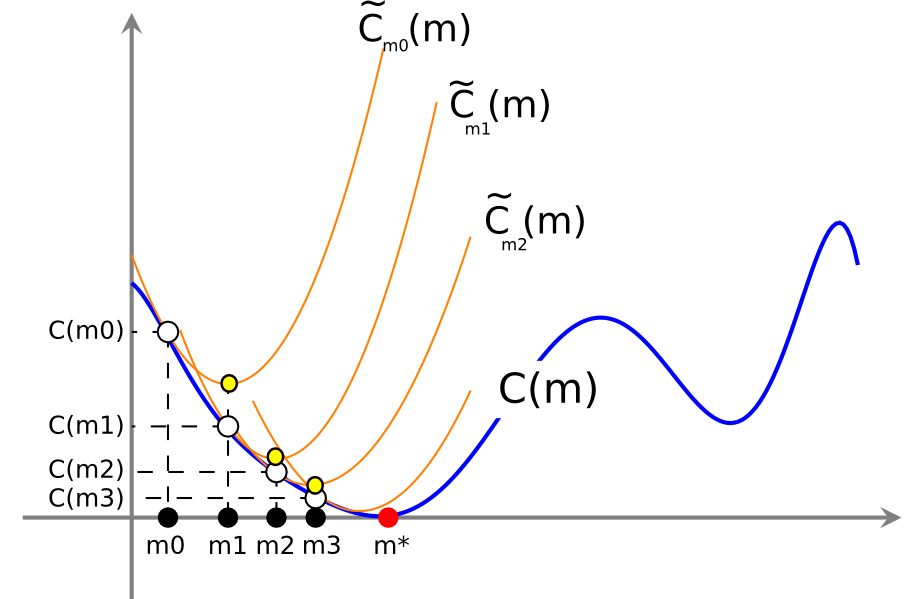} 
\caption{Sketch of the multimodal nonlinear FWI data fitting misfit function $C^{FWI}(\bold{m})$ (blue curve), equation~\ref{eqfwimf}, and the sequence of surrogate quadratic functions $\tilde{C}_{\bold{m}_k}(\bold{m})$ built at each iteration of Newton-type algorithms (orange curves) by linearization around the background models $\bold{m}_k$, where $\tilde{C}_{\bold{m}_k}(\bold{m}) = C(\bold{m}_k) + \delta \bold{m}^T \nabla_{\bold{m}} C(\bold{m}_k) +  \frac{1}{2} \delta \bold{m}^T \nabla^2_{\bold{m}} C(\bold{m}_k) \delta \bold{m}$ with $\bold{m}=\bold{m}_k+\delta \bold{m}$ \citep[e.g.,][]{Aghamiry_2020_FWI}.}
\label{fig_nonlinear}
\end{center}
\end{figure}

\noindent In the framework of local-optimization (gradient-based) methods, classical FWI recasts the non linear data fitting problem as a recurrence of surrogate linear problems, whose unknowns are the model perturbations with which the background model $\bold{m}$ of the current iteration will be updated and the observables are the data residuals $\delta \bold{d}^*(\bold{m})=\bold{d}^*-\bold{d}(\bold{m})$, where $\bold{d}(\bold{m})=\bold{S}(\bold{m})\bold{b}$ (Figure~\ref{fig_nonlinear}). In this framework, $\delta \bold{d}^*(\bold{m})$ should be understood as the recorded scattered data by the missing model perturbation $\delta \bold{m}^*$ in $\bold{m}$ \citep{Pratt_1998_GNF}. Accordingly, the data residuals are the restriction at receivers of the true scattered wavefields $\delta \bold{u}^*$ by $\delta \bold{m}^*$.
Applying the observation operator $\bold{P}$ to $\delta \bold{u}^*$ in equation~\ref{eqdu*} gives
\begin{equation}
\delta \bold{d}^*(\bold{m})  =  \bold{S}(\bold{m}) \delta \bold{b}^* =  -\omega^2 \bold{S}(\bold{m}) \sum_i w_i \boldsymbol{\delta}(\bold{x}-\bold{x}_i),
\label{eqrsd}
\end{equation}
\noindent where it is reminded that $w_i=\delta \bold{m}_i^* \bold{u}_i^*$. The ground-truth scattered data are then the recording at receivers of the blended impulse response of the background $\bold{m}$ over the region containing $\delta \bold{m}^*$ where the  unknown weights depend both on $\delta \bold{m}^*$ and $\bold{u}^*$.
The nonlinear Equation~\ref{eqrsd} can be rewritten as
\begin{equation}
\delta \bold{d}(\delta \bold{m})= \bold{B}(\bold{m};\bold{u}(\bold{m}+\delta \bold{m})) \delta \bold{m},
\end{equation}
where $\bold{B}(\bold{m};\bold{u}(\bold{m}+\delta \bold{m})) =  -\omega^2 \bold{S}(\bold{m}) \text{diag}(\bold{u}(\bold{m}+\delta \bold{m}))$.
\noindent Equation~\ref{eqrsd} can be viewed as a reparametrization of the nonlinear forward equation of FWI, $\bold{d}= \bold{S}(\bold{m})\bold{b}$.
In classical FWI, the simulated scattered data are computed after a linearization of this nonlinear forward equation in the frame of the Born approximation, namely $\bold{B}(\bold{m};\bold{u}(\bold{m})) \approx \bold{B}(\bold{m};\bold{u}(\bold{m}+\delta \bold{m}))$.
%

\noindent The gradient of FWI is formed by the inner product between each column of the sensitivity matrix and the data residuals  \citep[e.g.,][]{Pratt_1998_GNF}: 
\begin{equation}
\nabla_{\bold{m}} \phi^{FWI}(\bold{m}) = \bold{J}(\bold{m})^T \delta \bold{d}(\bold{m})^*,
\end{equation}
where
\begin{equation}
\bold{J}=
\left[
\begin{array}{cccccc}
\frac{\partial d_1}{\partial m_1} & \frac{\partial d_1}{\partial m_2} &  \cdot & \cdot &  \cdot & \frac{\partial d_1}{\partial m_M} \\
\frac{\partial d_2}{\partial m_1} & \frac{\partial d_2}{\partial m_2} &  \cdot & \cdot & \cdot & \frac{\partial d_2}{\partial m_M} \\
\cdot  & \cdot &  \cdot & \cdot & \cdot & \cdot  \\
\frac{\partial d_j}{\partial m_1}   & \frac{\partial d_j}{\partial m_2} &  \cdot & \cdot & \cdot & \frac{\partial d_j}{\partial m_M} \\
\cdot   & \cdot &  \cdot & \cdot & \cdot & \cdot \\
\frac{\partial d_N}{\partial m_1} & \frac{\partial d_N}{\partial m_2} &  \cdot & \cdot & \cdot & \frac{\partial d_N}{\partial m_M} 
\end{array}
\right]
\text{,} ~~
\delta \bold{d}(\bold{m})^* = \left[
\begin{array}{c}
\delta d_1(\bold{m}) \\
\delta d_2(\bold{m}) \\
\cdot    \\
\delta d_j(\bold{m}) \\
\cdot    \\
\delta d_N(\bold{m}) \\
\end{array}
\right].
\nonumber
\end{equation}
\noindent For one frequency, $M$ and $N = n_s \times n_r$ denote the number of grid points sampling $\bold{m}$ and the number of data sampling the stationary-recording acquisition, respectively.

\begin{figure}[htb!]
\centering
\begin{center}
\includegraphics[width=17cm,clip=true,trim=0cm 9cm 0.cm 0cm]{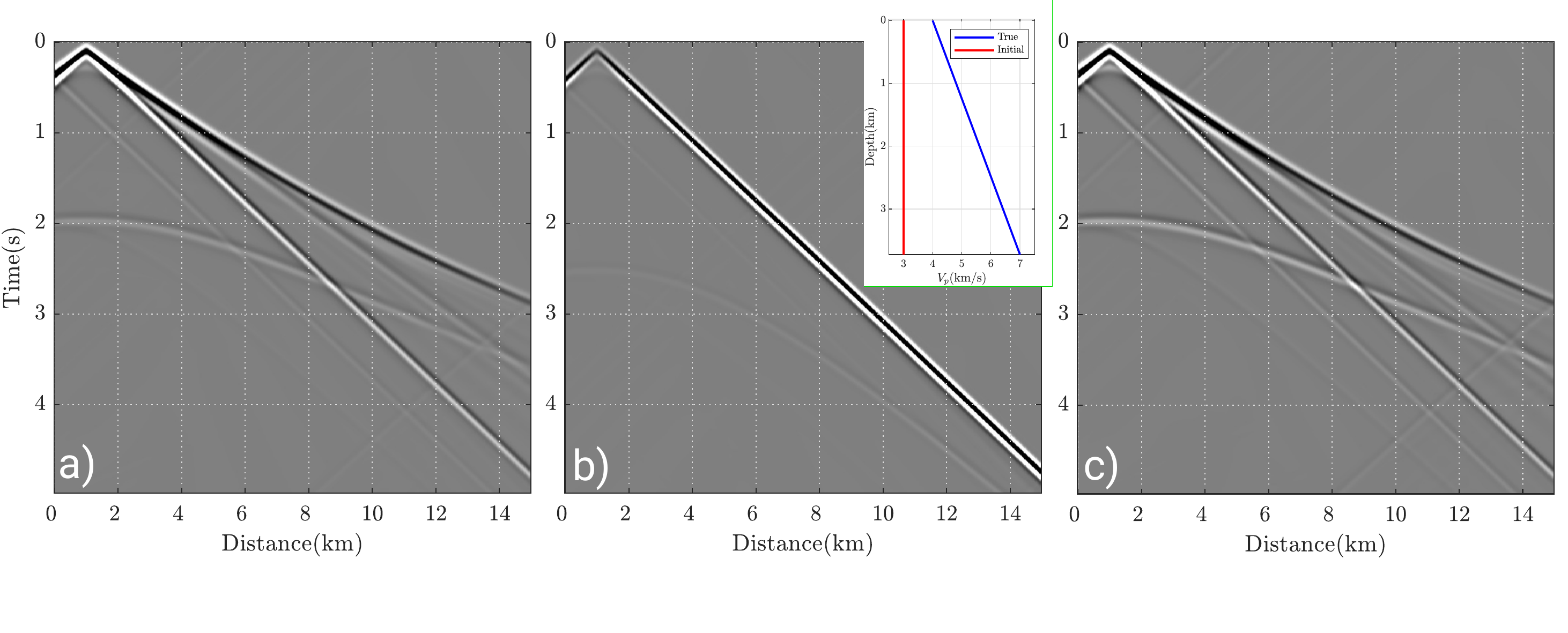} 
\caption{Scattered data and the Born approximation. Inset: The ground-truth model $\bold{m}^*$ (blue curve) is a vertical velocity-gradient model while the background medium $\bold{m}$ is homogeneous (red curve). (a) Ground-truth scattered data: $\delta \bold{d}^*(\bold{m})=\bold{d}^*-\bold{d}(\bold{m})=- \omega^2 \bold{S}(\bold{m}) \text{diag}(\bold{m}^*) \bold{u}^*$. (b) Scattered data in the Born approximation:  $\delta \bold{d}^{Born}(\bold{m})=- \omega^2 \bold{S}(\bold{m}) \text{diag}(\bold{m}^*) \bold{u}$. (c) Scattered data using $\bold{u}^e$ in the scattering source:  $\delta \bold{d}(\bold{m})=- \omega^2 \bold{S}(\bold{m}) \text{diag}(\bold{m}^*) \bold{u}^e$.}
\label{fig_Born}
\end{center}
\end{figure}

\noindent The simulated scattered data $\frac{\partial \bold{d}}{\partial m_i}$ are the restriction at receivers of the so-called partial derivative wavefields, which are the solution of an approximate Lippmann-Schwinger equation for a point source located at the position of the model parameter $i$ with respect to which the partial derivative is computed \citep[][ their equation 15]{Pratt_1998_GNF}. \\
\noindent For column $i$,
\begin{equation}
\partial_{\bold{m}_i} \bold{d}(\bold{m}) = -\omega^2 \bold{S}(\bold{m}) w_i^{FWI} \boldsymbol{\delta}(\bold{x}-\bold{x}_i),
\label{eqssd}
\end{equation}
\noindent where $w_i^{FWI} = \bold{u}_i$. Comparing equations~\ref{eqrsd} and \ref{eqssd} shows that the sensitivity matrix can be interpreted as a deblending operator whose columns aim at picking in the weights $w_i$ of the blended scattering source that model perturbation $\delta \bold{m}_i^*$ to be mapped at the position of  $\bold{m}_i$. The key difference between equations~\ref{eqrsd} and \ref{eqssd} lies indeed in the wavefield incident to the scatterers in their scattering sources (the true wavefield and the background wavefield in $w_i$ and $w_i^{FWI}$, respectively). This difference results from the weak-scattering Born approximation with which the simulated scattered data are computed as above mentioned. This approximation is poor when $\delta \bold{m}$ generates strong scattering along the incident path connecting the source to the scatterers and prevents accurate picking of the model perturbation by the Born-simulated scattered data as sketched in Figure~\ref{fig_lsequation} and illustrated numerically in Figure~\ref{fig_Born}.


\noindent According to the above review, we will show that the key contribution of ES-FWI is simply to implement more accurate wavefields  in the virtual source of the partial derivative data beyond the Born approximation.

\subsection{Estimating more accurate wavefields from the Lippmann-Schwinger equation}
\noindent We estimate the wavefields that best approximate $\bold{u}^*$ with the Lippmann-Schwinger equation because this equation gathers all the available priors about $\bold{u}^*$, that is the propagator $\bold{m}$ and the recorded data $\bold{d}^*$, i.e., the restriction of $\bold{u}^*$ at receivers. Let's remind that the true wavefield satisfies the wave equation
\begin{equation}
\bold{A}(\bold{m}) \bold{u}^* = \bold{b} +  \delta \bold{b}^*,
\label{equ*2}
\end{equation}
\noindent where $\delta \bold{b}^*$ is the unknown of the following rank-deficient forward equation
\begin{equation}
\delta \bold{d}^* = \bold{S}(\bold{m}) \delta \bold{b}^*.
\label{eqdddb}
\end{equation}
\noindent We can compute an approximation of $\bold{u}^*$ by solving the wave equation, equation~\ref{equ*2}, once we have estimated an approximation of $\delta \bold{b}^*$. The best we can do to estimate $\delta \bold{b}^*$ is to solve the rank-deficient equation \ref{eqdddb} in a least-squares sense

\begin{equation}
\delta \bold{b}^e = \text{arg} \min_{\delta \bold{b}} \| \delta \bold{d}^*(\bold{m})  - \bold{S}(\bold{m}) \delta \bold{b} \|_2^2 + \mu\| \delta \bold{b} \|_2^2,
\label{eqdb0}
\end{equation}
\noindent where a damping regularization was introduced for stabilization. This problem is nothing more than the damped least-squared scattered-data fitting problem for scattering source estimation.

\noindent The solution of this source-dependent underdetermined problem is given by
\begin{equation}
\delta \bold{b}^e(\bold{m}) = \bold{S}^T(\bold{m}) \left(\bold{S}(\bold{m}) \bold{S}^T(\bold{m}) + \mu \bold{I}\right)^{-1} \delta \bold{d}^* = \bold{S}^T(\bold{m}) \bold{H}_d^{-1} (\bold{m}) \delta \bold{d}^*,
\label{eqdbe1}
\end{equation}
\noindent where the Hessian is formulated in the data domain at the expense of the source domain since the dimension of the data space spanned by one source (here, $n_r$ when considering one frequency) is much smaller than the domain where the wavefields are computed. Moreover, let's note that $\delta \bold{b}^e$ is related to $\delta \bold{b}^*$ by the source resolution matrix $\bold{R}(\bold{m})$,
\begin{equation}
\delta \bold{b}^e = \left( \bold{S}(\bold{m})^T \bold{S}(\bold{m}) + \mu \bold{I} \right)^{-1} \bold{S}^T(\bold{m}) \delta \bold{d}^* = \left( \bold{S}(\bold{m})^T \bold{S}(\bold{m}) + \mu \bold{I} \right)^{-1} \bold{S}^T(\bold{m}) \bold{S}(\bold{m})  \delta \bold{b}^* = \bold{R}(\bold{m}) \delta \bold{b}^*,
\label{eqdbe2}
\end{equation}
and, the simulated scattered data $\delta \bold{d}^e$ by $\delta \bold{b}^e$ are related to the recorded scattered data $\delta \bold{d}^*$ by the data-domain resolution matrix $\bold{R}_d(\bold{m})$:
\begin{equation}
\delta \bold{d}^e = \bold{S}(\bold{m}) \delta \bold{b}^e = \bold{S}(\bold{m}) \bold{S}^T(\bold{m}) \left(\bold{S}(\bold{m}) \bold{S}^T(\bold{m}) + \mu \bold{I}\right)^{-1} \delta \bold{d}^* = \bold{R}_d(\bold{m}) \delta \bold{d}^*.
\label{eqdde}
\end{equation}
It is important to stress that the underdetermination of the scattering-source estimation not only results from the incomplete subsurface illumination provided by the receiver layout but more importantly by the fact that the sought $\delta \bold{b}^*$ depends on two classes of unknowns, $\delta \bold{m}^*$ and $\bold{u}^*$, with potential coupling between the two. \\
\noindent From the numerical viewpoint, equation~\ref{eqdbe1} shows that the scattering sources $\delta \bold{b}_s^e$ can be computed in two steps: \\

\noindent For $s=1,...,n_s$,
\begin{enumerate}

\item Solve approximately or exactly the normal system

\begin{equation}
\left(\bold{S}(\bold{m}) \bold{S}^T(\bold{m}) + \mu \bold{I}\right) \bold{y} = \delta \bold{d}_s^*.
\label{eqnormal}
\end{equation}

\item Solve the adjoint wave equation

\begin{equation}
\bold{A}^T(\bold{m})  \delta \bold{b}_s^e(\bold{m}) = \bold{P}^T \bold{y}
\label{eqadjoint}
\end{equation}
\noindent with the weighted data residuals as source. Note that the left operator in equation~\ref{eqnormal} is independent to the sources for a stationary-recording acquisition.

\end{enumerate}

\noindent The scattering sources $\delta \bold{b}_s^e$ show obvious analogy with the so-called adjoint wavefield of classical FWI, the difference being related to the weighting operator applied to the data residuals before back-propagation. The role of this weighting operator will be discussed in more detail later. 
Finally, one can readily see that the $\delta \bold{b}$ problem, equation~\ref{eqdb0}, is nothing more that a reparametrization of wavefield reconstruction problem, equation~\ref{penalty}.
\begin{equation}
\left[ \bold{u} \rightarrow \delta \bold{b} \right] \Rightarrow \left[ \min_{ \bold{u}} \| \bold{d}^*  - \bold{P} \bold{u} \|_2^2 + \mu\| \bold{A}(\bold{m}) \bold{u} -  \bold{b} \|_2^2 \rightarrow \min_{\delta \bold{b}} \| \delta \bold{d}^*  - \bold{S}(\bold{m}) \delta \bold{b} \|_2^2 + \mu\| \delta \bold{b} \|_2^2 \right].
\label{equwri}
\end{equation}
Therefore, 
the best-fitting wavefield $\bold{u}^e$  in the extended space can be inferred from the best-fitting scattering source $\delta \bold{b}^e$ by solving the wave equation in the background $\bold{m}$ with the following extended source $\bold{b}^e$:
\begin{equation}
\bold{A}(\bold{m})\bold{u}^e = \bold{b}^e = \bold{b} + \delta \bold{b}^e(\bold{m}) =\bold{b} + \bold{R}(\bold{m}) \delta \bold{b}^*(\bold{m}). 
\label{eque}
\end{equation}
\noindent Analogy between equations~\ref{equ*2} and \ref{eque} shows that $\delta \bold{b}^e$ accounts approximately for the scattering generated by the unknown $\delta \bold{m}^*$ contained in $\delta \bold{b}^*$, equation~\ref{eqdu*}, where the approximation is represented by the source-domain resolution matrix $\bold{R}(\bold{m})$. Figure~\ref{fig_Born}c shows that using $\bold{u}^e$ in the scattering source of the Lippmann-Schwinger equation in place of the background wavefield $\bold{u}(\bold{m})$ allows for the accurate modeling of the ground-truth scattered data in presence of strong scattering unlike the Born approximation (Figure~\ref{fig_Born}b).

\subsection{Parameter estimation beyond the Born approximation}
\noindent Now that we saw how to reconstruct wavefields that best match the true wavefields, equation~\ref{eque}, we review how they modify the gradient with respect to $\bold{m}$ of the classical FWI misfit function.
\noindent Injecting the expression of $\bold{u}^e$, equation~\ref{eque}, and $\delta \bold{b}^e$, equation~\ref{eqdbe1}, in the ES-FWI objective function, equation~\ref{penalty}, and reorganizing the terms recasts the later as a generalized form of the classical FWI misfit function \citep{vanLeeuwen_2019_ANO,Symes_2020_WRI,Gholami_2022_EFW}
\begin{equation}
\min_{ \bold{m}} \sum_{s=1}^{n_s} \| \delta \bold{d}_s^*(\bold{m}) \|_{\bold{H}_{d}^{-1} (\bold{m})}^2,
\label{equwri}
\end{equation}
\noindent where $\|\bold{y}\|_{\bold{Q}}^2=\bold{y}^T\bold{Q} \bold{y}$. \\
\noindent The data-domain Hessian $\bold{H}_d (\bold{m})$ acts as a data-domain covariance matrix, which however depends on $\bold{m}$.
\noindent Let's remind that 
\begin{equation} 
\mathcal{C}(\bold{m}) =  \| \delta \bold{d}(\bold{m}) \|_{\bold{Q}}^2 \rightarrow \nabla \mathcal{C}(\bold{m}) = \left[ \frac{ \partial \bold{d}(\bold{m}) }{\partial \bold{m} } \right]^T \bold{Q} \delta \bold{d}(\bold{m}).
\end{equation}
\noindent The above equation shows that introducing a covariance matrix in the FWI data misfit function modifies the adjoint source (i.e., the data residuals) in its gradient.
However, as $\bold{H}_d^{-1} (\bold{m})$ depends on $\bold{m}$, it not only modifies the source of the adjoint equation but also the kinematic of the simulated scattered data in the sensitivity matrix. This can be easily shown by computing the gradient of the ES-FWI objective function with respect to $\bold{m}$ by keeping $\bold{u}^e$ fixed and independent of $\bold{m}$. This amounts to minimize the scattering sources $\delta \bold{b}_s^e(\bold{m})=\bold{A}(\bold{m}) \bold{u}_s^e - \bold{b}_s$ in a least-squares sense:
\begin{equation}
\min_{\bold{m}} \sum_{s=1}^{n_s}\| \delta \bold{b}_s^e(\bold{m}) \|_2^2.
\label{eqmfdb}
\end{equation}
\textit{Remark 1}: We drop the penalty parameter $\mu$ as a multiplicative factor in the above equation because it is useless since the gradient has not the units of the parameters and hence a metric (step length and/or Hessian) should be defined accordingly to design the descent direction of the $\bold{m}$-subproblem (see also \textit{Remark 4}). \\

\noindent The gradient is given by
\begin{equation}
\nabla_\bold{m} \sum_{s=1}^{n_s}\| \delta \bold{b}_s^e(\bold{m}) \|_2^2  = \nabla_\bold{m} \sum_{s=1}^{n_s}\| \bold{A}(\bold{m}) \bold{u}_s^e - \bold{b}_s \|_2^2= \sum_{s=1}^{n_s} \left(\frac{\partial \bold{A}(\bold{m}) \bold{u}_s^e}{\partial \bold{m}}\right)^T \delta \bold{b}_s^e.
\label{eqgrad0}
\end{equation}
\noindent Substituting $\bold{u}_s^e$ by $\bold{u}_s + \bold{A}^{-1}(\bold{m}) \bold{R}_s(\bold{m}) \delta \bold{b}_s^*(\bold{m})$ and $\delta \bold{b}_s^e$ by $\bold{S}^T(\bold{m})  \bold{H}_{d}^{-1} (\bold{m}) \delta \bold{d}_s^*(\bold{m})$ in the above equation highlights the similarities and differences between FWI and ES-FWI gradients:
%
%
\begin{eqnarray}
\nabla_\bold{m}  \sum_{s=1}^{n_s} \| \delta \bold{b}_s^e(\bold{m}) \|_2^2 & = & \sum_{s=1}^{n_s} \left(\overbrace{\bold{S}(\bold{m}) \frac{\partial \bold{A}(\bold{m}) \bold{u}_s}{\partial \bold{m}} }^{single-scattering}\right)^T \bold{H}_{d}^{-1} (\bold{m}) \delta \bold{d}_s^* \nonumber \\
 & + & \sum_{s=1}^{n_s}  \left(\overbrace{\bold{S}(\bold{m})\frac{\partial \bold{A}(\bold{m})}{\partial \bold{m}}  \underbrace{ \bold{A}^{-1}(\bold{m}) \bold{R}_s(\bold{m}) \delta \bold{b}_s^*}_{\delta \bold{u}_s^e \approx \delta \bold{u}_s^*}}^{multi-scattering} \right)^T \bold{H}_{d}^{-1} (\bold{m}) \delta \bold{d}_s^*.
 \label{eqnablam}
\end{eqnarray}
\noindent The first term is the gradient of classical FWI except that the data residuals in the adjoint source are weighted by $\bold{H}_{d}^{-1} (\bold{m})$, the inverse of the data-domain Hessian of the $\delta \bold{b}$ estimation problem, equation~\ref{eqdb0}. The term in brackets represents the partial derivative data of FWI computed with the weak-scattering Born approximation. That is, the so-called virtual sources are built by sampling the incident background wavefields $\bold{u}_s$ to the scatterers with the radiation pattern matrix $\partial \bold{A}(\bold{m})/\partial \bold{m}$ from which the single-scattered partial derivative wavefields are propagated and recorded at receivers through the forward modeling operator $\bold{S}(\bold{m})$ \citep{Pratt_1998_GNF}. The second term supplements the single-scattered component of the partial derivative data with an approximation of the multi-scattered counterpart. This is shown by reading this term from right to left. An approximation $\delta \bold{u}_s^e$ of the true scattered wavefields $\delta \bold{u}_s^*$,
\begin{equation}
\delta \bold{u}_s^e =  \bold{A}^{-1}(\bold{m}) \delta \bold{b}_s^e = \bold{A}^{-1}(\bold{m}) \bold{R}(\bold{m}) \delta \bold{b}_s^*,
\label{eqdue5}
\end{equation}
\noindent are triggered by  $\delta \bold{b}_s^e=\bold{R}(\bold{m}) \delta \bold{b}_s^*$, i.e., the approximation of the true scattering sources $\delta \bold{b}_s^*$. These incident scattered wavefields are then sampled by the radiation pattern matrix $\partial \bold{A}(\bold{m})/\partial \bold{m}$ to build the multi-scattered component of the virtual sources from which the multi-scattered component of the partial derivative wavefields are propagated and recorded at receivers via the forward modeling operator $\bold{S}(\bold{m})$. The approximation with which scattering generated by the sought model perturbation is estimated in the incident wavefields is highlighted by the resolution matrix $\bold{R}(\bold{m})$ of the scattering-source estimation problem, equation~\ref{eqdbe2}. \\

\noindent This analogy between FWI and ES-FWI highlights that ES-FWI can be easily implemented in a classical FWI code although the extra ingredients of ES-FWI can generate significant albeit manageable computational overheads. These ingredients are the scattered wavefields added to the background wavefields in the virtual sources  and the inverse of the data-domain Hessian in the adjoint source. 
These two ingredients can be easily turned off in the code to switch from ES-FWI to FWI once the updated model lies in the basin of attraction of the global minimizer. \\

\textit{Remark 2}: The attentive reader will have noted that the estimation of the true scattering sources, namely $\delta \bold{b}_s^e$, are involved two times in the ES-FWI algorithm. First, they are used to supplement the background wavefield with the scattered wavefield by the missing model perturbation in the background model to build an estimation of the true wavefields, equation~\ref{eque}. 
These wavefields are then used to build more accurate virtual sources of the partial derivative data, term in brackets in equation~\ref{eqgrad0}. 
Second, $\delta \bold{b}_s^e$ are used as the so-called adjoint wavefields, i.e., the estimation of the true scattering sources, which are correlated with the simulated virtual sources to build the gradient of ES-FWI.


\subsection{Parameter estimation as a linear problem}
\noindent In the previous section, we formulate the parameter-estimation subproblem in a way that would highlight the similarities and differences between the steepest-descent direction of FWI and ES-FWI. Alternatively, when the wavefields and the model parameters are updated in alternating mode, the subsurface model can be obtained directly taking advantage of the bilinearity of the wave equation reviewed at the beginning of this tutorial. Assuming that the wavefields are known and independent of $\bold{m}$, the objective function of ES-FWI for model estimation can be recast as a quadratic function
\begin{equation}
\min_{ \bold{m}} \sum_{s=1}^{n_s}\| \bold{L}(\bold{u}^e_s) \bold{m} -  \bold{y}_s(\bold{u}^e_s) \|_2^2,
\label{equwri}
\end{equation}
\noindent where $\bold{L}(\bold{u}^e_s) = \omega^2 \text{diag}(\bold{u}_s^e)$ and $\bold{y}_s = \bold{b}_s - \nabla^2 \bold{u}_s^e$, equation~\ref{eqA2} \citep{VanLeeuwen_2013_MLM,Aghamiry_2019_IWR}. \\
\noindent The solution of this overdetermined problem reads
\begin{equation}
\bold{m} = \frac{\sum_{s=1}^{n_s} \bold{L}^T(\bold{u}^e_s) \bold{y}_s(\bold{u}^e_s)}{\sum_{s=1}^{n_s}  \bold{L}^T(\bold{u}^e_s)  \bold{L}(\bold{u}^e_s)},
\label{eqbilio}
\end{equation}
\noindent which is indeed the least-squares analogue of the expression of $\bold{m}$ inferred from the bilinearity of the wave equation when a monochromatic wavefield triggered by a single source is known everywhere in the subsurface, equation~\ref{eqbili}.
The denominator of the right-hand side of equation~\ref{eqbilio} shows that the Gauss-Newton Hessian is diagonal and is formed by the zero-lag autocorrelation of the virtual sources in the extended search space. It corrects the descent direction of ES-FWI from geometrical spreading along the incident paths connecting the sources to the scatterers \citep{Shin_2001_IAP}. The missing component of the Gauss-Newton FWI Hessian accounting for the paths connecting the scatterers to the receivers has been moved in the wavefield estimation subproblem through the data-domain Hessian of the scattering-source estimation problem. Disregarding the damping term, this Hessian is formed by the normal matrix $\bold{S}\bold{S}^T$ where $\bold{S}$ describes the propagation from the scatterers to the receivers. Therefore, $\bold{S}\bold{S}^T$ describes the summation over scatterers of the correlation in time and space between Green functions recorded at different receivers. Applying the inverse of this operator to the recorded scattered data aims at deconvolving the later from the limited bandwidth effects generated by these correlations \citep[][ Their appendix A]{Gholami_2022_EFW}. We will show however later that this deconvolution has not the same effect on the recorded data and the simulated data in the data residuals when the background $\bold{m}$ is highly inaccurate. \\

\subsection{Summary of the overall ES-FWI workflow}
\noindent As a summary, we recap the two step-procedure of ES-FWI.
The starting point is to recognize that the true wavefields satisfy a wave equation that can be written in two different forms:
\begin{eqnarray}
\bold{A}(\bold{m}^*) \bold{u}^* & = & \bold{b},   ~~~~~~~~~~~~~~~ \text{Classical wave equation} \\
\bold{A}(\bold{m})  \bold{u}^* & = & \bold{b} + \delta \bold{b}^*,  ~~~~~~~ \text{Lippmann-Schwinger volume integral equation}
\end{eqnarray}
\noindent where $\delta \bold{b}^*$ is the unknown of the rank-deficient forward-problem equation
\begin{equation}
\delta \bold{d}^* = \bold{S}(\bold{m}) \delta \bold{b}^*.
\label{eqddd}
\end{equation}

{\it{Wavefield reconstruction: Pushing $u$ towards $u^*$ with Lippmann-Schwinger equation}} \\

\noindent According to the Lippmann-Schwinger equation satisfied by $\bold{u}^*$ and $\bold{u}^e$,
\begin{eqnarray}
& & \bold{A}(\bold{m})  \bold{u}^*  =  \bold{b} + \delta \bold{b}^*,  \\
& & \bold{A}(\bold{m})  \bold{u}^e  =  \bold{b} + \delta \bold{b}^e, 
\end{eqnarray}
\noindent we seek $\bold{u}^e$ that best fits  $\bold{u}^*$ by looking for $\delta \bold{b}^e$ that best fits $\delta \bold{b}^*$.
\noindent From equation~\ref{eqddd}, $\delta \bold{b}^e$ is defined as the damped least-squares solution of the scattered-data fitting problem: \\
\noindent For $s=1,...,n_s$,
\begin{equation}
\delta \bold{b}_s^e = \text{arg} \min_{\delta \bold{b}_s} \| \delta \bold{d}_s^*(\bold{m})  - \bold{S}(\bold{m}) \delta \bold{b}_s \|_2^2 + \mu\| \delta \bold{b}_s \|_2^2,
\label{eqdb}
\end{equation}
which gives
\begin{equation}
\delta \bold{b}_s^e = \bold{S}^T(\bold{m}) \bold{H}_d^{-1} (\bold{m}) \delta \bold{d}_s^*(\bold{m}).
\label{eqdb10}
\end{equation}

{\it{Parameter estimation: Pushing $m$ towards $m^*$ by minimizing the scattering sources}}

\noindent According to
\begin{eqnarray}
\bold{A}(\bold{m}^*) \bold{u}^* & = & \bold{b} + \bold{0}, \label{eqm1} \\
\bold{A}(\bold{m})  \bold{u}^e & = & \bold{b} + \delta \bold{b}^e, \label{eqm2}
\end{eqnarray}
\noindent and assuming that $\bold{u}^e$ is a good approximation of $\bold{u}^*$, pushing $\bold{m}$ towards $\bold{m}^*$ amounts to jointly minimize $\delta \bold{b}^e$ for each source $\bold{b}$ of the experiment while minimizing the residual data misfit:

\begin{equation}
\min_{ \bold{m}} \sum_{s=1}^{n_s} \| \bold{d}_s^*(\bold{m})  - \bold{P} \bold{u}_s^e(\bold{m}) \|_2^2 + \mu \sum_{s=1}^{n_s} \| \bold{A}(\bold{m}) \bold{u}_s^e(\bold{m}) -  \bold{b}_s \|_2^2,
\label{equwri}
\end{equation}
\noindent which is equivalent to
\begin{equation}
\min_{ \bold{m}} \sum_{s=1}^{n_s} \| \delta \bold{d}_s^*(\bold{m}) \|_{\bold{H}_d^{-1} (\bold{m})}^2.
\label{equwrivp}
\end{equation}
\noindent If $\bold{u}_s^e$ and $\bold{m}$ are updated in alternating mode rather than through a variable projection method, the $\bold{m}$-subproblem reduces to the minimization of the scattering sources (adjoint wavefields),
\begin{equation}
\min_{\bold{m}} \sum_{s=1}^{n_s}\| \delta \bold{b}_s^e(\bold{m}) \|_2^2,
\label{equwriad}
\end{equation}
\noindent which is quadratic in $\bold{m}$ by virtue of the bilinarity of the wave equation, equation~\ref{equwri}, and amounts to minimize the data residuals $\delta \bold{d}^*(\bold{m})$ according to equation~\ref{eqdb10}. \\
\noindent The gradient of the two objective functions, equations~\ref{equwrivp} and \ref{equwriad}, are the same, while the Hessian is not \citep{vanLeeuwen_2016_PMP,Gholami_2022_EFW}. 

\subsection{On the accuracy of the reconstructed wavefields in extended search space}
Understanding the limits of ES-FWI and designing relevant heuristics to drive it toward accurate minimizer require additional insights on the accuracy with which the reconstructed wavefields match the true counterparts.
Let's first remind the expression of $\bold{u}^e$ from equations~\ref{eque} and \ref{eqdbe1}:\\
For  $s=1,...,n_s$,
\begin{equation}
\bold{A}(\bold{m}) \bold{u}_s^e = \bold{b}_s^e(\bold{m}) = \bold{b}_s + \bold{S}^T(\bold{m}) \left(\bold{S}(\bold{m}) \bold{S}^T(\bold{m}) + \mu \bold{I}\right)^{-1} \delta \bold{d}_s^*(\bold{m})
\label{eque3}
\end{equation}
and let's denote the simulated data in the extended space by $\bold{d}_s^e = \bold{P} \bold{u}_s^e$. \\
Reordering the terms of equation~\ref{eque3} provides the following identities: \\
For  $s=1,...,n_s$,
\begin{eqnarray}
\bold{b}_s^e(\bold{m}) & = &  (\bold{I} - \bold{R}(\bold{m}) ) \bold{b}_s +  \bold{S}(\bold{m})^T \bold{H}_d^{-1}(\bold{m}) \bold{d}_s^*,                                                                  \label{id1} \\
\bold{u}_s^e(\bold{m}) & = &  \bold{A}^{-1}(\bold{m}) (\bold{I} - \bold{R}(\bold{m}) ) \bold{b}_s + \bold{A}^{-1}(\bold{m}) \bold{A}(\bold{m})^{-T} \bold{P}^T \bold{H}_d^{-1}(\bold{m}) \bold{d}_s^*,      \label{id2} \\
\bold{d}_s^e(\bold{m}) & = & (\bold{I} - \bold{R}_d(\bold{m})) \bold{d}_s(\bold{m}) +  \bold{R}_d(\bold{m}) \bold{d}_s^*,                   \label{id3} \\
\Delta \bold{d}_s^e(\bold{m}) & = & \mu \left( \bold{S}(\bold{m}) \bold{S}^T(\bold{m}) + \mu \bold{I}\right)^{-1} \delta \bold{d}_s^*(\bold{m})=\left(\bold{I}-\bold{R}_d(\bold{m}) \right) \delta \bold{d}^*,                                                           \label{id4}
\end{eqnarray}
where $\Delta \bold{d}_s^e(\bold{m}) = \bold{d}_s^* - \bold{d}_s^e(\bold{m})$ are the data residuals in the extended search space, which should not be confused with the simulated scattered data in the extended search space, $\delta \bold{d}_s^e =  \bold{d}_s^e(\bold{m}) - \bold{d}(\bold{m}) = \bold{S}(\bold{m}) \delta \bold{b}_s^e =  \bold{R}_d(\bold{m}) \delta \bold{d}^*$, equation~\ref{eqdde}. 

\begin{figure}[htb!]
\centering
\begin{center}
\includegraphics[width=17cm,clip=true,trim=0cm 0cm 0.cm 0cm]{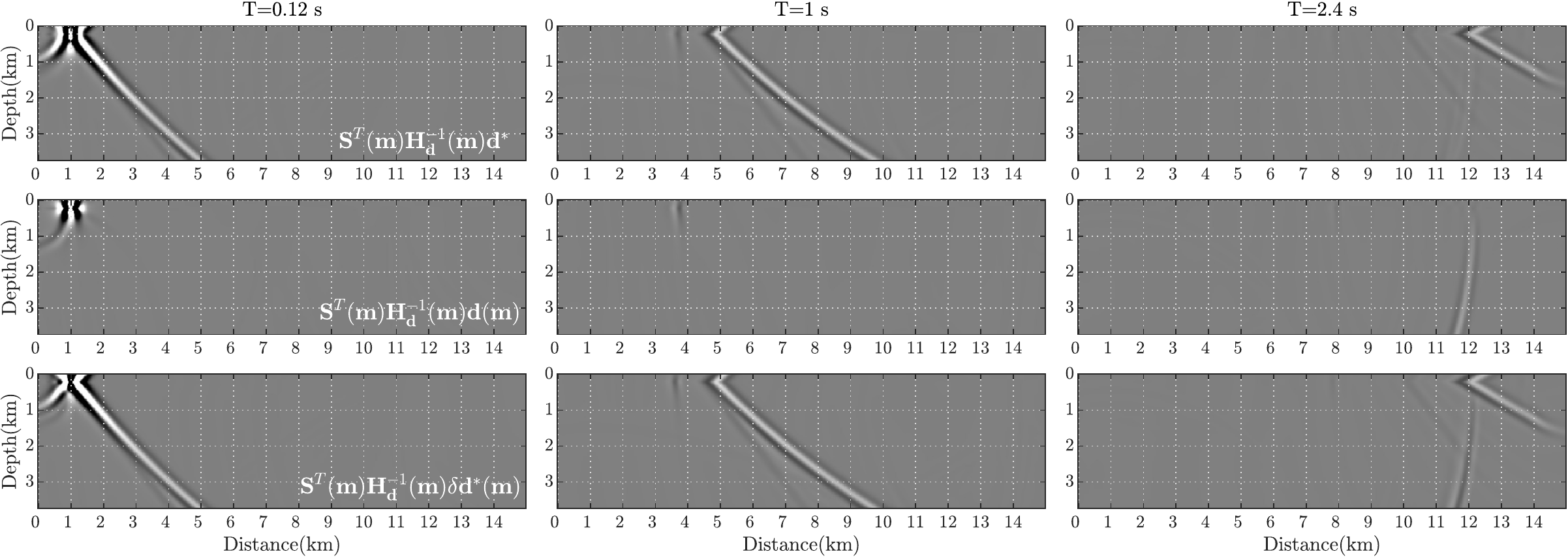} 
\caption{Scattering source. Snapshots of (a) the weighted back-propagated recorded data, (b) the weighted back-propagated simulated data, and (c) the weighted back-propagated data residuals.}
\label{fig_be}
\end{center}
\end{figure}

\noindent The extended sources $\bold{b}_s^e(\bold{m})$, equation~\ref{id1}, are the weighted sum of $\bold{b}_s$ and the back-propagated weighted recorded data $\bold{H}_d^{-1}(\bold{m}) \bold{d}_s^*$. When $\mu$ is set to a small value, the source resolution matrix $\bold{R} \approx \bold{I}$ and $\bold{b}_s^e(\bold{m})$ mostly correspond to the back propagated weighted recorded data. The reconstruction of the extended source $\bold{b}^e$ is illustrated in Figure~\ref{fig_be} for the velocity models shown in Figure~\ref{fig_Born}. Figure~\ref{fig_be}b shows the back propagated simulated data. Due to the kinematic consistency of the simulated data $\bold{d}(\bold{m})$ with the background $\bold{m}$, the back-propagated simulated data optimally match the point source $\bold{b}$ in space and time. Conversely, the back-propagated recorded data are smeared in time and space as the witness of the inaccuracy of $\bold{m}$ (Figure~\ref{fig_be}a). When the weighted data residuals are back-propagated and summed with $\bold{b}$, the sign-reversed simulated component of the back-propagated data residuals cancels out with $\bold{b}$ by subtraction mostly leaving the back propagated recorded data in $\bold{b}^e$ (Figure~\ref{fig_be}c).

\textit{Remark 3}: The back-propagation of the recorded data to estimate the source in the extended search space is more directly highlighted when ES-FWI is parametrized with the source rather than with the wavefield or the scattering source \citep[][ their equations 10 and 11]{Huang_2018_SEW}. However, the connection with inverse scattering theory and hence classical FWI is less obvious with the source parametrization.


\begin{figure}[htb!]
\centering
\begin{center}
\includegraphics[width=17cm,clip=true,trim=0cm 9cm 0.cm 0cm]{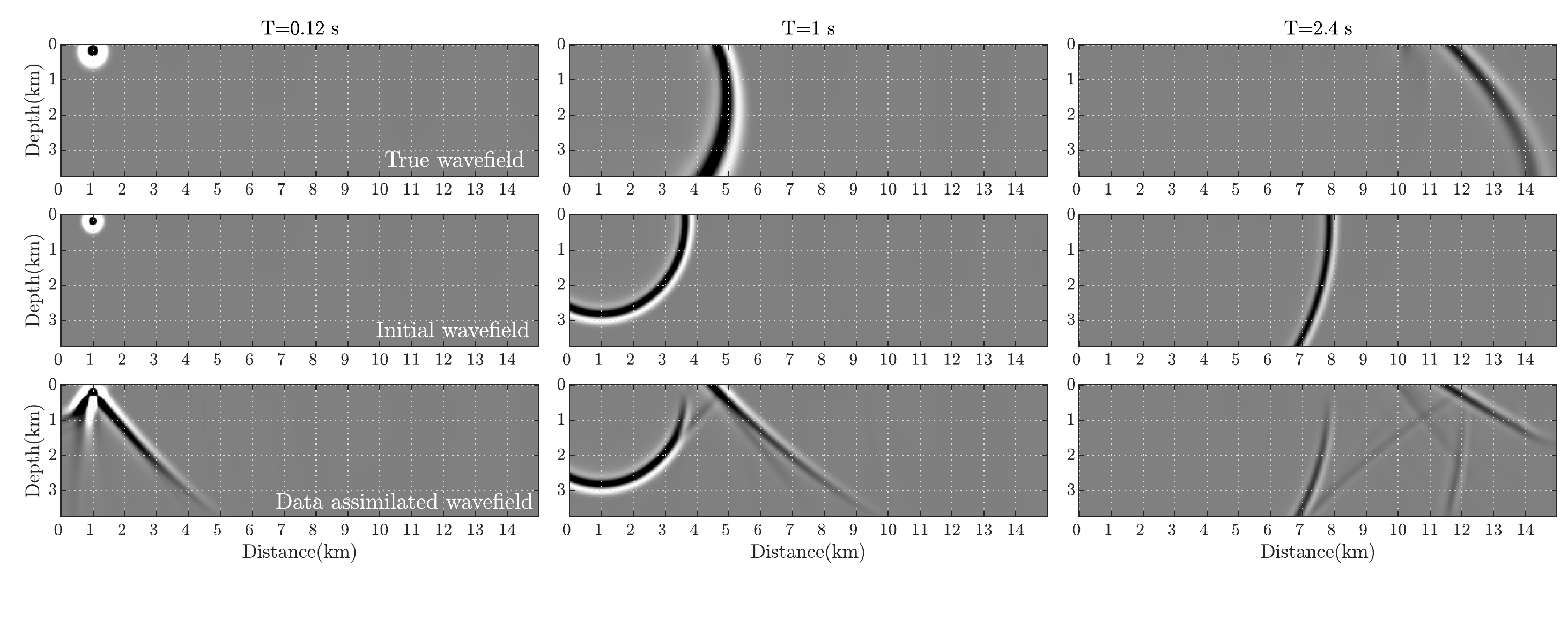} 
\caption{Data-assimilated wavefield. Snapshots of (a) the true wavefield $\bold{u}^*$, (b) the background wavefield $\bold{u}$, and (c) the data-assimilated wavefield $\bold{u}^e$. }
\label{fig_ue}
\end{center}
\end{figure}

\noindent From the expression of the extended sources, the extended wavefields $\bold{u}_s^e$, equation~\ref{id2}, can be formulated as the weighted sum of the background wavefields and the migrated-demigrated weighted recorded data (the migration-demigration operator is given by $\bold{A}^{-1}(\bold{m}) \bold{A}(\bold{m})^{-T}\bold{P}^T$). When $\mu$ is set to a small value, $\bold{R}_d(\bold{m}) \approx \bold{I}$ and the extended wavefields mostly reduce to the migrated-demigrated weighted recorded data. Accordingly, they match the recorded data at receivers by virtue of wave invariance for time reversal. However, they accumulate kinematic errors as they propagate away from the receivers in the background $\bold{m}$. 
This is illustrated in Figure~\ref{fig_ue}, which shows snapshots of the true wavefield $\bold{u}^*$, the background wavefield $\bold{u}$ and the data-assimilated wavefield $\bold{u}^e$ for the velocity models of Figure~\ref{fig_Born}. One can see that $\bold{u}^e$ matches $\bold{u}^*$ only near the receivers when $\bold{m}$ differs significantly from $\bold{m}^*$ while it fits $\bold{u}$ in the deep part with a complex transition between the shallow to the deep regions.


\begin{figure}[htb!]
\centering
\begin{center}
\includegraphics[width=17cm,clip=true,trim=0cm 12cm 0.cm 0cm]{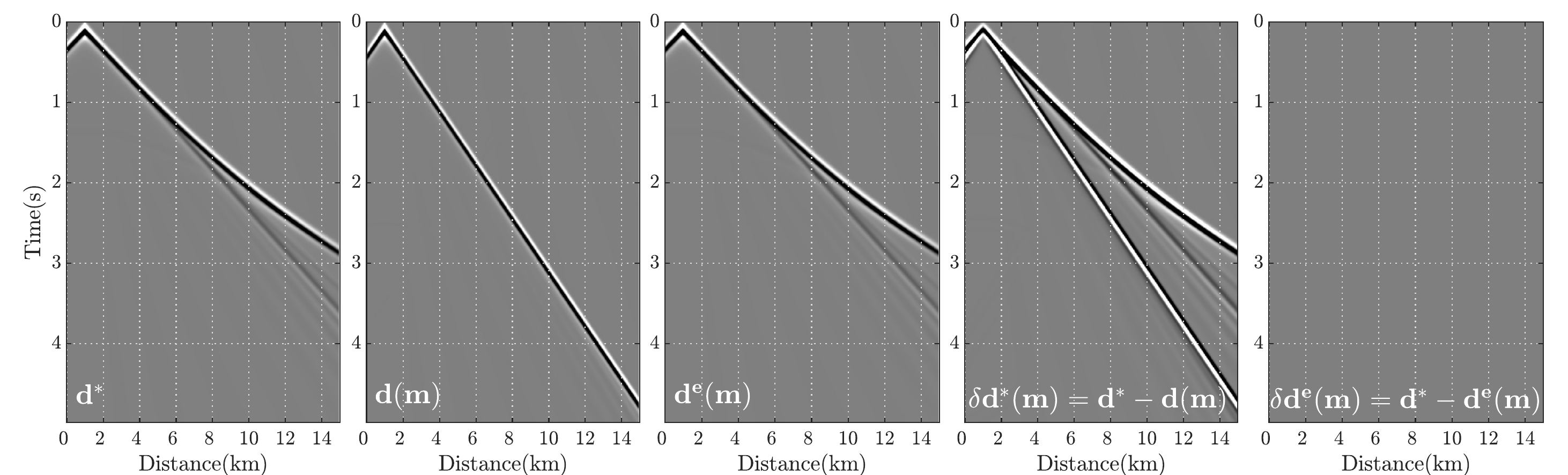} 
\caption{(a) Recorded data. (b) Simulated data in the reduced parameter space, $\bold{d}(\bold{m}) = \bold{S}(\bold{m}) \bold{b}$. (c) Simulated data in the extended space, equation~\ref{id3}, as the convex combination of the recorded data and simulated data. (d) Data residuals in the reduced parameter space (the recorded scattered data). (e) Data residuals in the extended space.}
\label{fig_de}
\end{center}
\end{figure}

\noindent From the decomposition of $\bold{u}_s^e$, equation~\ref{id2}, the simulated data $\bold{d}_s^e$ in the extended space are the convex combination of the simulated data in $\bold{m}$, $\bold{d}_s(\bold{m}) = \bold{S}(\bold{m}) \bold{b}_s$, and the recorded data $\bold{d}_s^*$ where the weighting operator is the data-resolution matrix. Again, setting $\mu$ to a small value pushes $\bold{d}^e$ towards  $\bold{d}_s^*$ at the expense of $\bold{d}_s(\bold{m})$ to prevent cycle skipping. The ability to match the recorded data from arbitrary background model $\bold{m}$ is further checked numerically in Figure~\ref{fig_de} with the velocity models shown in Figure~\ref{fig_Born}. The idea to push the simulated data towards the recorded data \citep[e.g.,][]{Baek_2014_RLS} or the reverse \citep[e.g.,][]{Yao_2019_TCS} is shared by almost all the approaches aiming at avoiding cycle skipping with the distinctiveness of ES-FWI to be physics driven.
 
\noindent Finally, equation~\ref{id4} shows that the data residuals in the extended space are linearly related to the recorded scattered data via the data-domain Hessian of the $\delta \bold{b}$ subproblem and further scaled by the penalty parameter $\mu$. This scaling highlights how the amplitudes of $\Delta \bold{d}_s^e$ decrease (i.e., how $\bold{d}_s^e$ can arbitrarily fit the recorded data) as $\mu$ decreases (Figure~\ref{fig_de}). The effect of the data-domain Hessian on the data residuals is discussed in the next section.

\textit{Remark 4}: The fact that $\bold{d}_s^e$  closely match $\bold{d}_s^*$ via the migration-demigration of the FWI data residuals, hence $\Delta \bold{d}_s^e$ are small, doesn't indeed mean that $\delta \bold{b}_s^e$, namely the scattering sources of the $\bold{u}^e$-subproblem and the adjoint wavefields of the $\bold{m}$-subproblem, are small. The sources of the adjoint equation satisfied by $\delta \bold{b}_s^e$  are the weighted data residuals of classical FWI by the inverse of the data domain Hessian, equation~\ref{eqdbe1}, that is $\left( 1 / \mu \right) \Delta \bold{d}_s^e$ according to equation~\ref{id4}. The division of $\Delta \bold{d}_s^e$ by the small $\mu$ restores the true strength of the scattering source. Similarly, the adjoint source of ES-FWI in the gradient of the ES-FWI misfit function is $\bold{H}_d^{-1}(\bold{m})^{-1} \delta \bold{d}^*$ and not $\Delta \bold{d}^e$ as highlighted by the generalized form of the ES-FWI misfit function, equation~\ref{equwri}. This can be further checked by computing explicitly the Gauss-Newton descent direction of the $\bold{m}$-subproblem for $\bold{u}^e$ fixed and independent of $\bold{m}$. \\
\noindent The gradient of the penalty objective function, equation~\ref{penalty}, is given by 
\begin{eqnarray}
\nabla_\bold{m} C^{P}(\bold{u}^e_s,\bold{m})|_{\bold{u}_s^e} & = & \mu \sum_s \left(\frac{\partial \bold{A}(\bold{m}) \bold{u}_s^e}{\partial \bold{m}} \right)^T \delta \bold{b}_s^e \\
& = &\mu \sum_s \left(\frac{\partial \bold{A}(\bold{m}) \bold{u}_s^e}{\partial \bold{m}} \right)^T \bold{S}^T(\bold{m}) \bold{H}_d^{-1}(\bold{m}) \delta \bold{d}_s^*(\bold{m}),
\end{eqnarray}

and the Gauss-Newton Hessian reads

\begin{equation}
\nabla_\bold{m}^2 C^{P}(\bold{u}^e_s,\bold{m})|_{\bold{u}_s^e} = \mu \sum_s \left(\frac{\partial \bold{A}(\bold{m}) \bold{u}_s^e}{\partial \bold{m}} \right)^T\left(\frac{\partial \bold{A}(\bold{m}) \bold{u}_s^e}{\partial \bold{m}} \right).
\end{equation}
Hence, the Gauss-Newton descent direction $\bold{p}_k$ is given by
\begin{equation}
\bold{p}_k(\bold{m})|_{\bold{u}_s^e} = \left( \nabla_\bold{m}^2 C^{P}(\bold{u}^e_s,\bold{m})|_{\bold{u}_s^e}\right)^{-1} \nabla_\bold{m} C^{P}(\bold{u}^e_s,\bold{m})|_{\bold{u}_s^e} = \frac {\sum_s \left(\frac{\partial \bold{A}(\bold{m}) \bold{u}_s^e}{\partial \bold{m}} \right)^T \bold{S}^T(\bold{m}) \bold{H}_d^{-1}(\bold{m}) \delta \bold{d}^*(\bold{m})}{\sum_s \left(\frac{\partial \bold{A}(\bold{m}) \bold{u}_s^e}{\partial \bold{m}}\right)^T\left(\frac{\partial \bold{A}(\bold{m}) \bold{u}_s^e}{\partial \bold{m}}\right)}.
\end{equation}
Note how $\mu$ is removed by division of the gradient by the Hessian.
Remembering the expressions of $\bold{u}_s^e$, equation~\ref{eque}, and $\delta \bold{b}^e_s$, equation~\ref{eqdbe1}, one can readily see that the penalty parameter $\mu$ is involved in the descent direction of the $\bold{m}$-subproblem only as a damping regularization term of the data-domain Hessian of the scattering source estimation problem consistently with the expression of the generalized FWI misfit function, equation~\ref{equwri}.


\subsection{On the role of the data-domain Hessian in the adjoint source}
\noindent It remains to us to discuss the role of the data-domain Hessian in the adjoint source of ES-FWI in the light of the former analysis of the data-assimilated wavefields $\bold{u}_s^e$. An heuristic interpretation is given hereafter. Let's first remind that this data-domain Hessian is the Hessian of a source location problem. Accordingly, let's imagine that we want to estimate the source $\bold{b}$ either from the pair $(\bold{d}^*,\bold{m}^*)$ or $(\bold{d}(\bold{m}),\bold{m})$ by solving the damped least-squares data-fitting problems
\begin{eqnarray}
&& \min_{\bold{b}} \| \bold{S}(\bold{m}) \bold{b} - \bold{d}(\bold{m})\|_2^2 + \mu \| \bold{b} \|_2^2, \\
&& \min_{\bold{b}} \| \bold{S}(\bold{m}^*) \bold{b} - \bold{d}^*)\|_2^2 + \mu \| \bold{b} \|_2^2.
\end{eqnarray}
\noindent The minimizers are
\begin{eqnarray}
\bold{b} = & \bold{S}^T(\bold{m}) \left(\bold{S}(\bold{m}) \bold{S}^T(\bold{m}) + \mu \bold{I} \right)^{-1} \bold{d}(\bold{m}) \\
\bold{b} = & \bold{S}^T(\bold{m}^*) \left(\bold{S}(\bold{m}^*) \bold{S}^T(\bold{m}^*) + \mu \bold{I} \right)^{-1} \bold{d}^*.
\end{eqnarray}
\noindent We can expect the two reconstructed sources to be similar as well as the effect of the data-domain Hessian on $\bold{d}(\bold{m})$ and $\bold{d}^*$ because the data are kinematically consistent with the model in both cases although a more focused reconstruction of $\bold{b}$ is expected from that model providing the best illumination of its wavenumber spectrum (for example, if $\bold{m}^*$ contains reflectors as opposed to a smooth $\bold{m}$, $\bold{b}$ may be better reconstructed from $\bold{m}^*$ due to the improved wavenumber illumination provided by back-scattered waves). Now, let's imagine that we wish to locate $\bold{b}$ from the recorded data and the background $\bold{m}$
\begin{equation}
\min_{\bold{b}} \| \bold{S}(\bold{m}) \bold{b} - \bold{d}^*\|_2^2 + \mu \| \bold{b} \|_2^2.
\end{equation}
\noindent The reconstructed source will be smeared in space and time due to the kinematic inconsistency between $\bold{d}^*$ and $\bold{m}$ and the focusing effect of the inverse of the data-domain Hessian $\bold{S}(\bold{m}) \bold{S}^T(\bold{m})$ on $\bold{d}^*$ may be less effective accordingly. 


\begin{figure}[htb!]
\centering
\begin{center}
\includegraphics[width=17cm,clip=true,trim=0cm 0cm 0.cm 0cm]{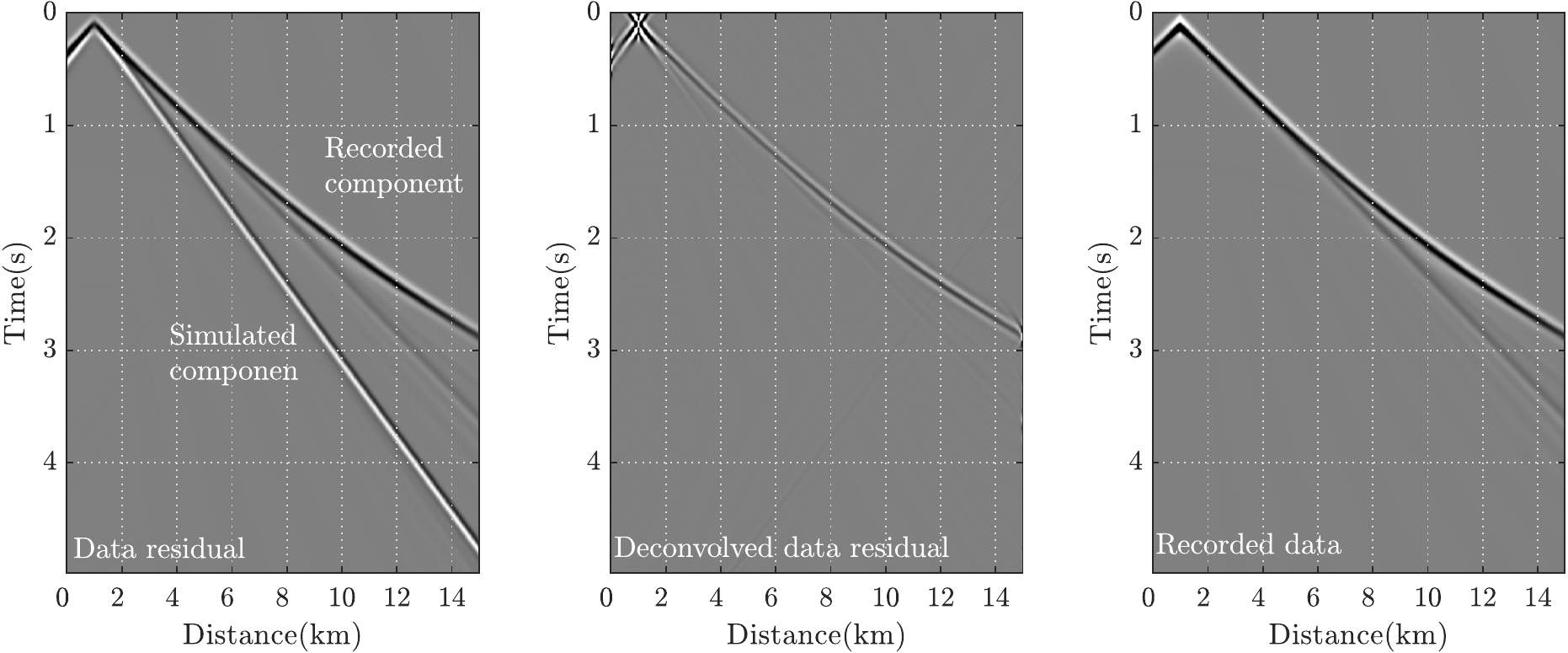} 
\caption{Effect of the data-domain Hessian on the data residuals. (a) Data residuals $\delta \bold{d}^*(\bold{m})$ for the velocity models of Figure~\ref{fig_Born}. (b) Data residuals after deconvolution by $\bold{H}_d^{-1}(\bold{m})$. (c) Recorded data $\bold{d}^*$.}
\label{fig_deltad}
\end{center}
\end{figure}

\noindent The contrasted effect of the data-domain Hessian on $\bold{d}(\bold{m})$ and $\bold{d}^*$ is shown in Figure~\ref{fig_deltad} for the velocity models shown in Figure~\ref{fig_Born}. The direct arrival in the background model has been focused at short offsets and near onset times such that the adjoint operator $\bold{S}^T(\bold{m})$ in equation \ref{eqdbe1} optimally focuses the back-propagated simulated data at the position of $\bold{b}$ and at onset times. Conversely, the data-domain Hessian has less visible effects on the recorded data. Therefore, we may conclude that the descent direction of the ES-FWI iterations is mostly driven by $\bold{d}^*$ at the expense of $\bold{d}(\bold{m})$ when $\bold{m}$ is strongly inaccurate due to the kinematic consistency of the incident wavefields $\bold{u}^e$ in the virtual sources of the simulated scattered data and the back-propagated recorded data, while the opposite behavior occurs in classical FWI due to the kinematic consistency between the incident background wavefield and the back-propagated simulated data.
The action of the data-domain Hessian in the adjoint source as a damping  of the simulated component of the data residuals accounts for the fact that the true scattering sources are modeled approximately in the sensitivity matrix of ES-FWI as highlighted by the resolution matrix $\bold{R}(\bold{m})$, equation~\ref{eqdbe1}-\ref{eqdbe2}. Accordingly, the simulated scattered data $\delta \bold{d}^e$ (and the partial derivative data) cannot predict well enough the recorded scattered data to capture the full differential information they carry out as highlighted by the data resolution matrix $\bold{R}_d(\bold{m})$ in equation~\ref{eqdde}. This is accounted for by weighting the data residuals in the adjoint source by the inverse of the Hessian of the scattering source location problem. This contrasts with the adjoint source of classical FWI where the data residuals are not weighted. This results because the weak-scattering Born approximation implicitly assumes that the simulated scattered data in the sensitivity matrix can capture the full differential information contained in the data residuals generated by weak scattering. Therefore, as ES-FWI converges toward an accurate $\bold{m}$, the scattering generated by $\delta \bold{m}^*$ becomes weaker and weaker, the effect of the data-domain Hessian on the simulated and recorded data becomes similar, and ES-FWI tends asymptotically toward FWI.


\section{Few remarks on algorithmic aspects}
\noindent Discussing in depth algorithmic aspects of ES-FWI is beyond the scope of this tutorial, which is focused on the physical principles governing ES-FWI. Here, we only briefly review for sake of completeness the algorithmic strategies that were addressed in previous studies or that will deserve further investigations to design workable ES-FWI algorithms with the best trade-off between computational efficiency and solution accuracy.

\subsection{Optimization algorithm and regularization}

\subsubsection{Why Augmented Lagrangian method?}
\noindent As mentioned in the section \textit{FWI as a constrained optimization problem: full space versus reduced space methods}, ES-FWI can be implemented with penalty method or augmented Lagrangian method. It is well acknowledged that one issue of penalty methods is related to the adaptive tuning of the penalty parameter such that a sufficient constraint relaxation is generated during the early iterations while the constraint is satisfied at the convergence point \citep{Fu_2017_DPM}. Augmented Lagrangian method allows for a constant penalty parameter to be used because the Lagrange multipliers will record the history of the constraint errors in iterations to progressively remove their footprint in iterations following an iterative refinement procedure (i.e., the iterative solution of an ill-posed linear problem)  \citep[][ Their Appendix A]{Aghamiry_2019_IWR}. As an illustration of this recording, one can readily check from equation~\ref{al} that the Lagrange multipliers $\bold{v}_s$, also referred to as the dual variables, reduce to the running sum of the constraint violations in iterations, when [i] the primal variables, namely $\bold{u}_s$ and $\bold{m}$, and the dual variables $\bold{v}_s$ are updated in alternating mode, and [ii] $\bold{v}_s$ are updated with basic gradient ascent steps. The reader is referred to \citet[][Chapter 17]{Nocedal_2006_NO} for a comparative analysis of penalty methods and augmented Lagrangian methods from a mathematical perspective, and \citet{Aghamiry_2019_IWR} for their assessment in the context of ES-FWI. 

\subsubsection{Implementing regularization with the alternating-direction method of multipliers (ADMM)}
\noindent ES-FWI is an underdetermined optimization problem as highlighted by the ill-posedness of the scattering source estimation problem, which requires some regularization. Another advantage of the augmented Lagrangian method (or method of multipliers) is to provide a versatile framework to implement various kinds of regularization including nonsmooth (non differential) ones with the alternating-direction method of multipliers (ADMM) \citep{Goldstein_2009_SBM,Esser_2018_TVR}. In this approach, some auxiliary variables are introduced in the regularized optimization problem via an additional constraint to decouple the least-squares problem from the $\ell{1}$-norm problem and recast the later as a denoising problem, which can be solved efficiently with proximal algorithms \citep{Parikh_2013_PA}. The resulting constrained problem is solved with ADMM by updating the various classes of primal variables in alternating mode. Moreover, the primal and the dual variables are also updated in alternating mode in the frame of the method of multipliers.
The reader is referred to \citet{Aghamiry_2019_IBC,Aghamiry_2019_CRO,Aghamiry_2020_FWI} for the implementation of various kinds of regularization in ES-FWI and FWI with ADMM.

\subsection{ES-FWI in the time domain}
\noindent ES-FWI was originally formulated in the frequency domain because it was unclear how to compute the data-assimilated wavefields with explicit time stepping schemes since these wavefields are the solution of an overdetermined linear system, \citep[e.g.][Their equation 6]{VanLeeuwen_2013_MLM} and hence of a normal equation \citep[e.g.][ Their equation 4]{Aghamiry_2019_AEW}. This tutorial has reviewed how to tackle this issue by solving the wave equation in the background model $\bold{m}$ with an extended source in the right hand side formed by the physical source plus a scattering source, equation~\ref{eque} and equations~\ref{eqnormal}-\ref{eqadjoint}. Another issue is related to the management of the Lagrange multipliers in the time domain since they have the size of the wavefields. \citet{Gholami_2022_EFW} showed how to project these Lagrange multipliers in the data domain through the relationship between $\delta \bold{b}^e$ and $\bold{H}_d^{-1} \delta \bold{d}^*$ via the adjoint of the wave-equation operator. Finally, compared to penalty methods, implementing ES-FWI with the augmented Lagrangian method just requires adding to the weighted data residuals $\bold{H}_d^{-1} \delta \bold{d}^*$ their running sum in iterations in the adjoint source of ES-FWI.

\subsection{Inversion preconditioning}
\noindent We have shown how the accuracy of the reconstructed wavefields decreases away from the receivers. It remains unclear under which conditions this varying accuracy will drive ES-FWI toward accurate or spurious subsurface models. To illustrate this point, inaccurate reconstructed wavefields sounds inconsistent with the linearization of the $\bold{m}$-subproblem around these wavefields. Therefore, one may investigate depth continuation or layer stripping approaches to further extend the linear regime of ES-FWI by updating first the shallow part of the subsurface before considering the deeper part. This might be implemented with appropriate covariance matrices in the data domain and/or in the source domain \citep{Lee_2020_SFW}.

\subsection{Waveform inversion workflow}
\noindent We have shown how ES-FWI evolves toward FWI as the inversion approaches the convergence point. Accordingly, one may view to degrade the accuracy with which some ingredients of ES-FWI are estimated during the late iterations to mitigate its computational burden or  switch to FWI during the latest stages of the inversion. On this subject, the data-domain Hessian of the scattering-source reconstruction subproblem generates the main computational overhead compared to classical FWI when its effects are taken into account accurately. Therefore, selecting the most suitable low-rank approximation of this data domain Hessian in iterations (diagonal approximation \citep{Gholami_2022_EFW}, matching filters \citep{Guo_2022_PID}, truncated Gauss-Newton approximation \citep{Guo_2022_PID}, point-spread function \citep{Lin_2022_TWR}, receiver encoding \citep{Aghamiry_2022_HWR}) to find the best compromise between computational cost and quality of the solution is one of the main issue of ES-FWI.


\section{Numerical examples}

\noindent We conclude this tutorial with two numerical illustrations corresponding to the Camembert model and the 2004 BP salt model. We would like to stress that the results obtained hereafter have been obtained with the augmented-Lagrangian implementation of ES-FWI. It would be very challenging if not impossible to obtain such results with the penalty method implementation even when adaptive tuning of the penalty parameter is used. Before showing the results of this test, we briefly discuss with a numerical example the convexity of the ES-FWI objective function.

\subsection{On the convexity of the ES-FWI objective function}
\begin{figure}[htb!]
\centering
\begin{center}
\includegraphics[width=16cm,clip=true,trim=0cm 8cm 0cm 8cm]{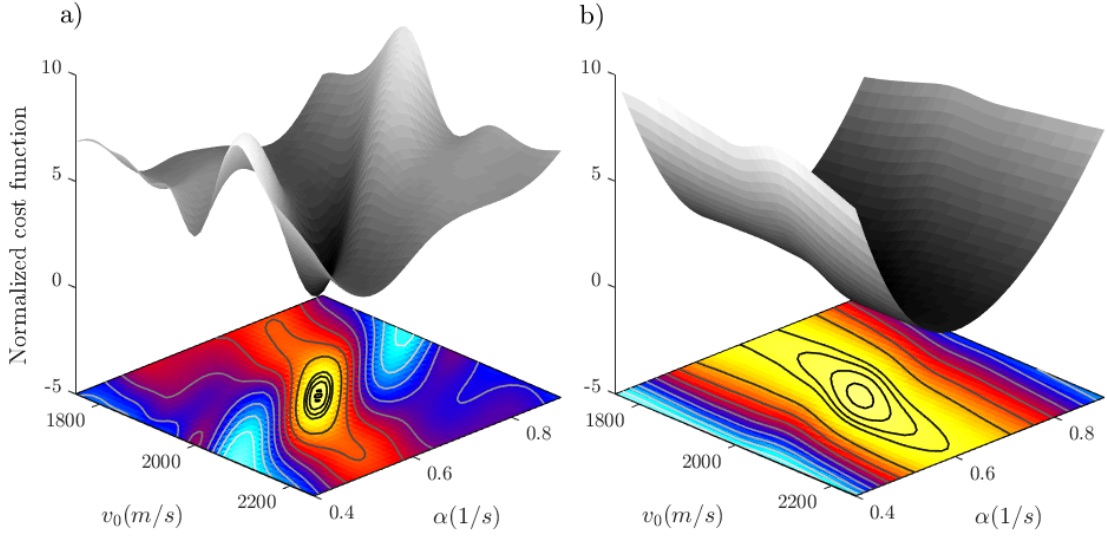} 
\caption{(a) FWI and (b) ES-FWI objective functions for the velocity gradient test.}
\label{fig_cost}
\end{center}
\end{figure}

To the best of our knowledge, there is no mathematical proof of the convexity of the ES-FWI objective function. Anyway, it is easy to design a numerical experiment that would trap ES-FWI in a spurious minimum suggesting that this convexity cannot be proved in situations met in exploration geophysics. The best we can propose is to illustrate with numerical examples that the objective function of ES-FWI is more convex than that of classical FWI. 
To do so, we consider a 2D laterally homogeneous model of horizontal and vertical dimensions $8~km \times 3~km$ where the velocity linearly increases with depth, $v(z)= v_0 + \alpha z$ \citep{Mulder_2008_ESI}. The acquisition consists of a single source located at x = 200~m, and a line of receivers spaced 50~m apart along the surface. 
The data are generated with a velocity v0 = 2000 m/s and a gradient $\alpha = 0.65~s^{-1}$. 
The cost function of FWI and ES-FWI when $\mu = 1e-2$ are computed in a  $41 \times 41$ grid covering the 2D search space spanned by the two parameters $v0$ and $\alpha$: $v0 \in [1750, 2250]$ and $alpha \in [0.4, 0.9]$ with steps of 12.5~m/s and 0.0125~1/s for v0 and $\alpha$, respectively. 
The FWI and ES-FWI cost functions are shown in Figure~\ref{fig_cost}. As expected the FWI cost function shows several local minima unlike the ES-FWI cost function. A similar test is shown in \citet{VanLeeuwen_2013_MLM} where $v0$ and $\alpha$ are processed separately.

%

\subsection{The Camembert test}

\begin{figure}[htb!]
\centering
\begin{center}
\includegraphics[width=16cm,clip=true,trim=0cm 0cm 0cm 0cm]{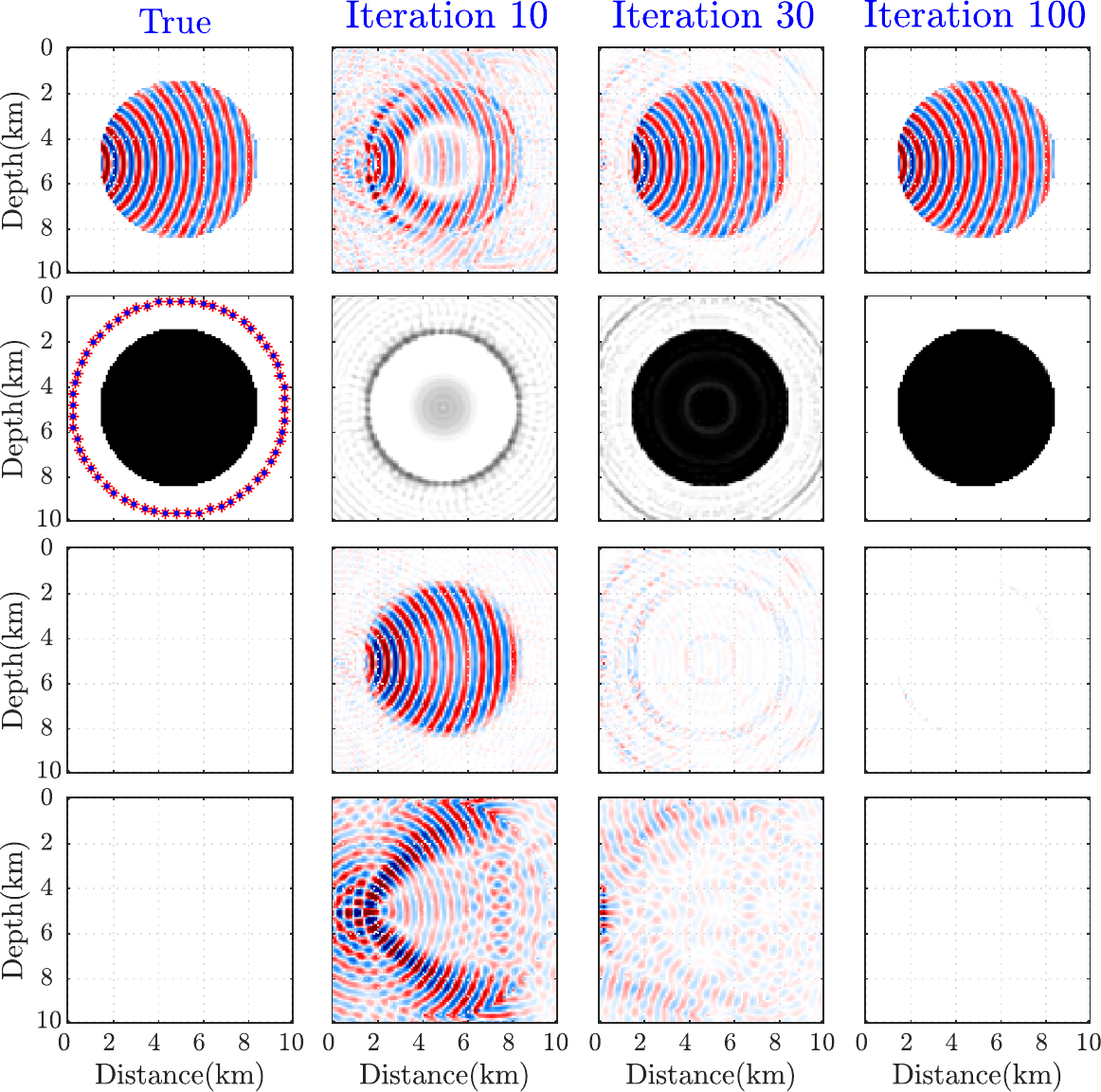} 
\caption{Camembert test. Top row: from left to right, $\delta \bold{b}(\bold{m};\bold{m}_0)$ for $\bold{m}=\bold{m}^*,\bold{m}^e_{k=10,30,100}$ as an alternative representation of the ground-truth wavefield $\bold{u}^*$ and the simulated wavefields in $\bold{m}^e_{k=10,30,100}$. Second row: From left to right, true model $\bold{m}^*$ and ES-FWI models $\bold{m}^e_{k=10,30,100}$. Third and fourth rows: Sought scattering sources $\delta \bold{b}(\bold{m}^*;\bold{m})$ and estimated scattering source $\delta \bold{b}^e(\bold{m})$ by ES-FWI, equation~\ref{eqdbe1}, for $\bold{m}=\bold{m}^*,\bold{m}_{10},\bold{m}_{30}, \bold{m}_{100}$.}
\label{fig_camembert}
\end{center}
\end{figure}

\noindent The Camembert model (Figure~\ref{fig_camembert}) contains a large circular inclusion in an homogeneous background \citep{Gauthier_1986_TDN}. The wavespeeds in the homogeneous background and in the inclusion are $V_0=3.2~km/s$ and $V_{inc}=3.5~km/s$, respectively, and the radius of the inclusion is 3.5~km.
A circular acquisition formed by 60 sources and receivers surrounds the inclusion to provide a full scattering-angle illumination of the target. Compared to the experimental setup designed by \citet{Gauthier_1986_TDN}, we use an ideal acquisition to focus on the resilience of ES-FWI to cycle skipping without acquisition footprint generated by incomplete illumination. We perform a mono-frequency inversion for the 5~Hz frequency to precisely quantify the amount of cycle skipping. Moreover, we apply  total-variation regularization to manage the ill-posedness of the scattering-source estimation. The starting model is the homogeneous background model ($V_0=3.2~km/s$).  Ten wavelengths are propagated in the inclusion($N_\lambda = 10$) and the strength of the contrast $\chi=(V_{inc}-V_0)/V_0$ is 20$\%$. The traveltime mismatch $\delta t$ accumulated during the propagation across the inclusion is 0.1875~s, which is almost two times greater than half the period (0.1~s). \\
\noindent The path followed by ES-FWI is illustrated in Figure~\ref{fig_camembert} at iterations 10, 30, 100. The top row illustrates, for the left most source, the true wavefield (left panel) and the simulated wavefields in the estimated models $\bold{m}^e_k$ by ES-FWI, namely $\bold{u}(\bold{m}^e_k) = \bold{A}^{-1}(\bold{m}_k^e) \bold{b}$. We show these wavefields in the equivalent form of their exact scattering source $\delta \bold{b}(\bold{m};\bold{m}_b)=-\omega^2 \text{diag}(\bold{m}-\bold{m}_{b})  \bold{u}(\bold{m})$ when the background model $\bold{m}_{b}$ is set to $\bold{m}_0$, namely
\begin{eqnarray}
\bold{u}^* & = &  \bold{A}^{-1}(\bold{m}^*) \bold{b} = \bold{A}^{-1} (\bold{m}_0) \left( \bold{b} + \delta \bold{b}(\bold{m}^*;\bold{m}_0) \right), \\
\bold{u}(\bold{m}^e_k) & = & \bold{A}^{-1}(\bold{m}^e_k) \bold{b} = \bold{A}^{-1} (\bold{m}_0) \left( \bold{b}  + \delta \bold{b}(\bold{m}_k^e;\bold{m}_0) \right).
\end{eqnarray}
The \textit{exact} scattering source depends on the full medium and the background medium and should not be confused with the \textit{estimated} scattering source from the background medium and the recorded data $\bold{d}^*$, equation~\ref{eqdbe1}. Accordingly, $\bold{u}(\bold{m}^e_k)$ should not be confused with the estimated $\bold{u}_k^e$ by ES-FWI from $\bold{m}_{k-1}^e$ and $\bold{d}^*$, equation~\ref{eque}.
\noindent The representation of the reconstructed wavefields by their scattering source provides a more direct assessment of their accuracy as manifested by the smearing of $\delta \bold{b}(\bold{m}^e_k;\bold{m}_0)$ compared to $\delta \bold{b}(\bold{m}^*;\bold{m}_0)$, the spatial support of the later matching that of the true inclusion (left panel). We see that $\delta \bold{b}(\bold{m}^e_{100};\bold{m}_{0})$ matches fairly well $\delta \bold{b}(\bold{m}^*;\bold{m}_{0})$ at the convergence point, which correlates with the good match between $\bold{m}^e_{100}$ and $\bold{m}^*$ (second row) according to the bilinearity of the wave equation. We also see that $\delta \bold{b}(\bold{m}^e_{10};\bold{m}_{0})$ still lacks the interior part of the true scattering source, which corresponds to the area away from the receivers. This correlates with the fact that $\bold{m}^e_{10}$ mostly contains the boundary of the inclusion (second row). The scenario where the short wavelengths are reconstructed before the longer ones (as opposed to the multiscale approach from long to short wavelengths implemented in classical FWI by frequency, time and offset continuations \citep[e.g.][]{Gorszczyk_2017_TRW}) generally traps FWI in spurious minima \citep[][ Their Figure 9]{Gauthier_1986_TDN}. We note however that $\delta \bold{b}(\bold{m}^e_{10};\bold{m}_{0})$ matches $\delta \bold{b}(\bold{m}^*;\bold{m}_0)$ also at receivers located far away from the source, hence highlighting the resilience of ES-FWI to cycle skipping. 
The reconstructed models in iterations (second row) show how the inclusion is progressively reconstructed from its boundary (near the receivers) to its center (far away from the receivers). Transposing this statement to surface acquisition would lead to the conclusion that ES-FWI reconstructs the subsurface from the shallow parts to the deeper parts as the accuracy of the wavefields improved away from receivers in iterations following some kinds of layer stripping.
The two bottom rows show the sought scattering sources $\delta \bold{b}(\bold{m}^*;\bold{m})$ and the estimated scattering sources $\delta \bold{b}^e(\bold{m})$ by ES-FWI  for $\bold{m}=\bold{m}^*, \bold{m}_{k=10, 30, 100}^e$. Indeed, the left panels  ($\bold{m}=\bold{m}^*$) are nil since the contrast are zero and hence the scattering source and the data residuals are zero too. 
The third row shows how $\delta \bold{b}(\bold{m}^*;\bold{m})$ vanishes as $\bold{m}$ tends to $\bold{m}^*$. Moreover, comparing the third and fourth rows highlights how the relative error between $\delta \bold{b}(\bold{m}^*;\bold{m}^e_k)$ and $\delta \bold{b}^e(\bold{m}^e_k)$ decreases as $\bold{m}^e_k$ tends to $\bold{m}^*$, i.e., as $\bold{u}_k^e$ tends to $\bold{u}^*$. This trend highlights how the ill-posedness of the scattering source reconstruction decreases as the estimated wavefield becomes closer to the true one, remembering that these scattering sources depend on two classes of unknowns, namely the sought contrast and the sought wavefield.
To summarize, the two-step workflow of ES-FWI (section \textit{Summary of the overall ES-FWI workflow}) is illustrated by noting how the reconstructed model and the wavefield reconstruction improve at the same time while the reconstructed scattering source vanishes. The central issue of ES-FWI is to mitigate the ill-posedness of the scattering source estimation problem due to the potential trade-off between the wavefields and the model perturbations.

%

\subsection{The 2004 BP salt benchmark}
\begin{figure}[htb!]
\centering
\begin{center}
\includegraphics[width=16cm,clip=true,trim=0cm 0cm 0cm 0cm]{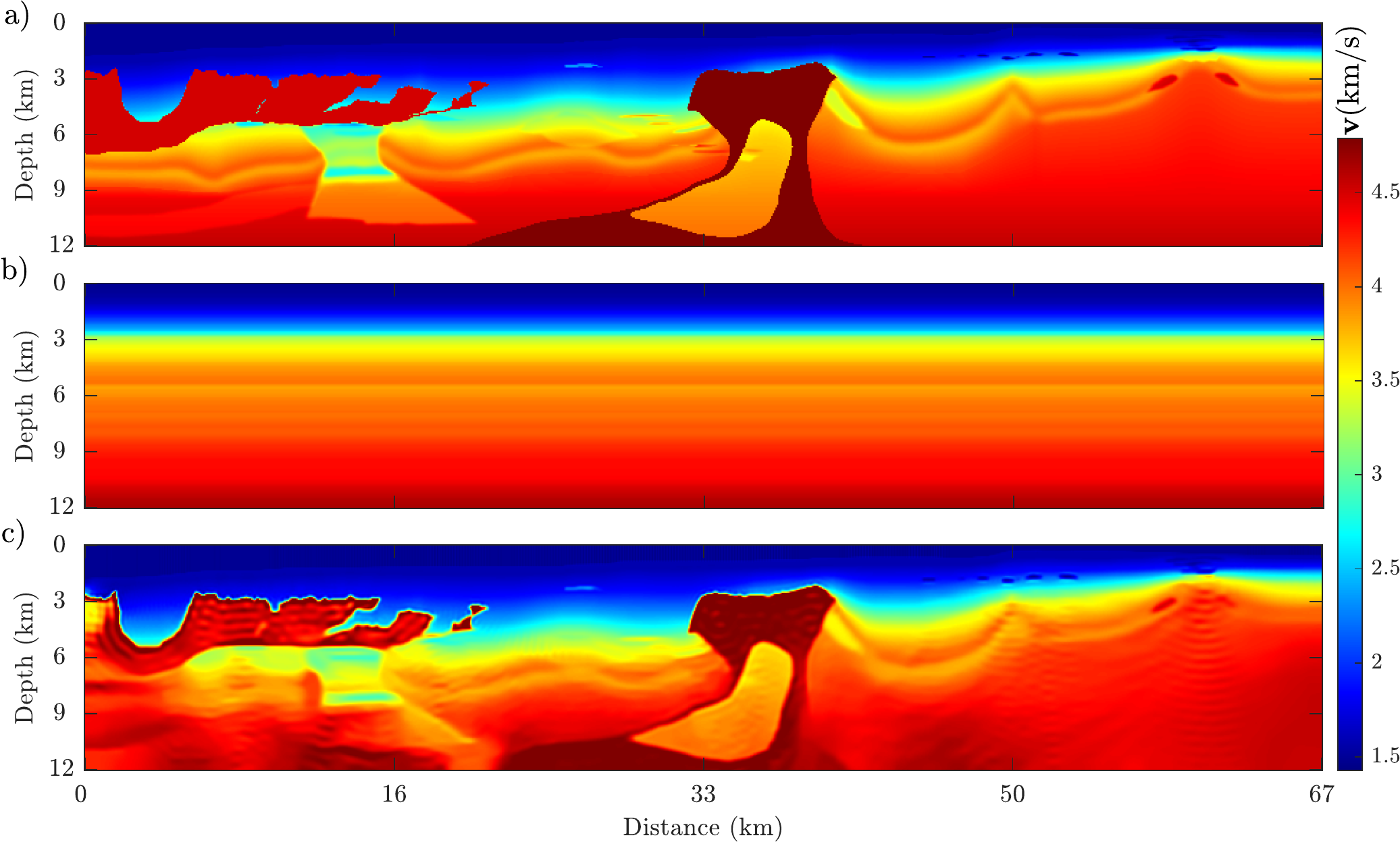} 
\caption{2004 BP salt model benchmark. (a) Ground-truth model $\bold{m}^*$. (b) Starting model $\bold{m}_0$. (c) Final ES-FWI model $\bold{m}^{ES-FWI}$.}
\label{fig_bpsalt}
\end{center}
\end{figure}

\noindent We conclude with the challenging 2004 BP salt model to illustrate the potential of ES-FWI for imaging complex subsurface models from inaccurate (smooth) starting model and ultra long-offset data when salt bodies generate a quite uneven illumination of the subsalt structure (Figure~\ref{fig_bpsalt}a).

\begin{figure}[htb!]
\centering
\begin{center}
\includegraphics[width=14cm,clip=true,trim=5cm 7cm 5cm 4cm]{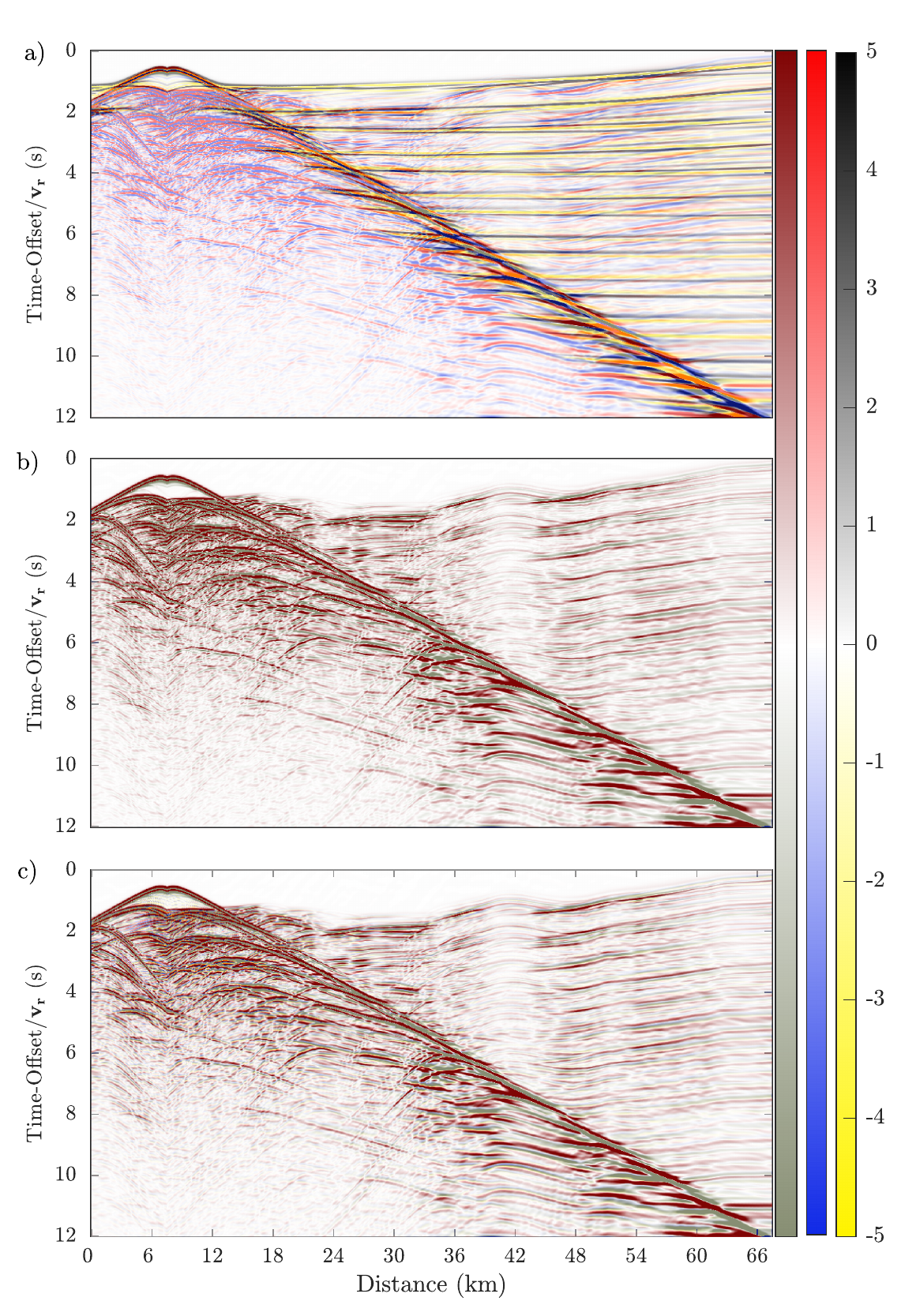} 
\caption{2004 BP salt model benchmark. (a-c) Direct comparison between (a) $\bold{d}^*=\bold{S}(\bold{m}^*) \bold{b}$ and $\bold{d}(\bold{m}_0)=\bold{S}(\bold{m}_0) \bold{b}$; (b)  $\bold{d}^*$ and $\bold{d}^e(\bold{m}_0)$; (c) $\bold{d}^*$ and $\bold{d}(\bold{m}_f^{e})=\bold{S}(\bold{m}_f^{e}) \bold{b}$, where $\bold{m}_f^{e}$ denotes the final model found by ES-FWI (Figure~\ref{fig_bpsalt}). In (a-c),  $\bold{d}(\bold{m}_0)$, $\bold{d}^e(\bold{m}_0)$ and  $\bold{d}(\bold{m}_f^{e})$ are superimposed in transparency on $\bold{d}^*$. The seismograms $\bold{d}(\bold{m}_0)$, $\bold{d}^e(\bold{m}_0)$ and  $\bold{d}(\bold{m}_f^{e})$ are plotted with black-white-yellow color scale (right scale) while $\bold{d}^*$ are plotted with a red-white-blue color scale (middle scale). The two sets of seismograms match when their superimposition appears as brown-white-olive colored seismograms (left scale), because black+red=brown and yellow+blue=olive.}
\label{fig_sismos}
\end{center}
\end{figure}

\begin{figure}[htb!]
\centering
\begin{center}
\includegraphics[width=16cm,clip=true,trim=0cm 3cm 0cm 0cm]{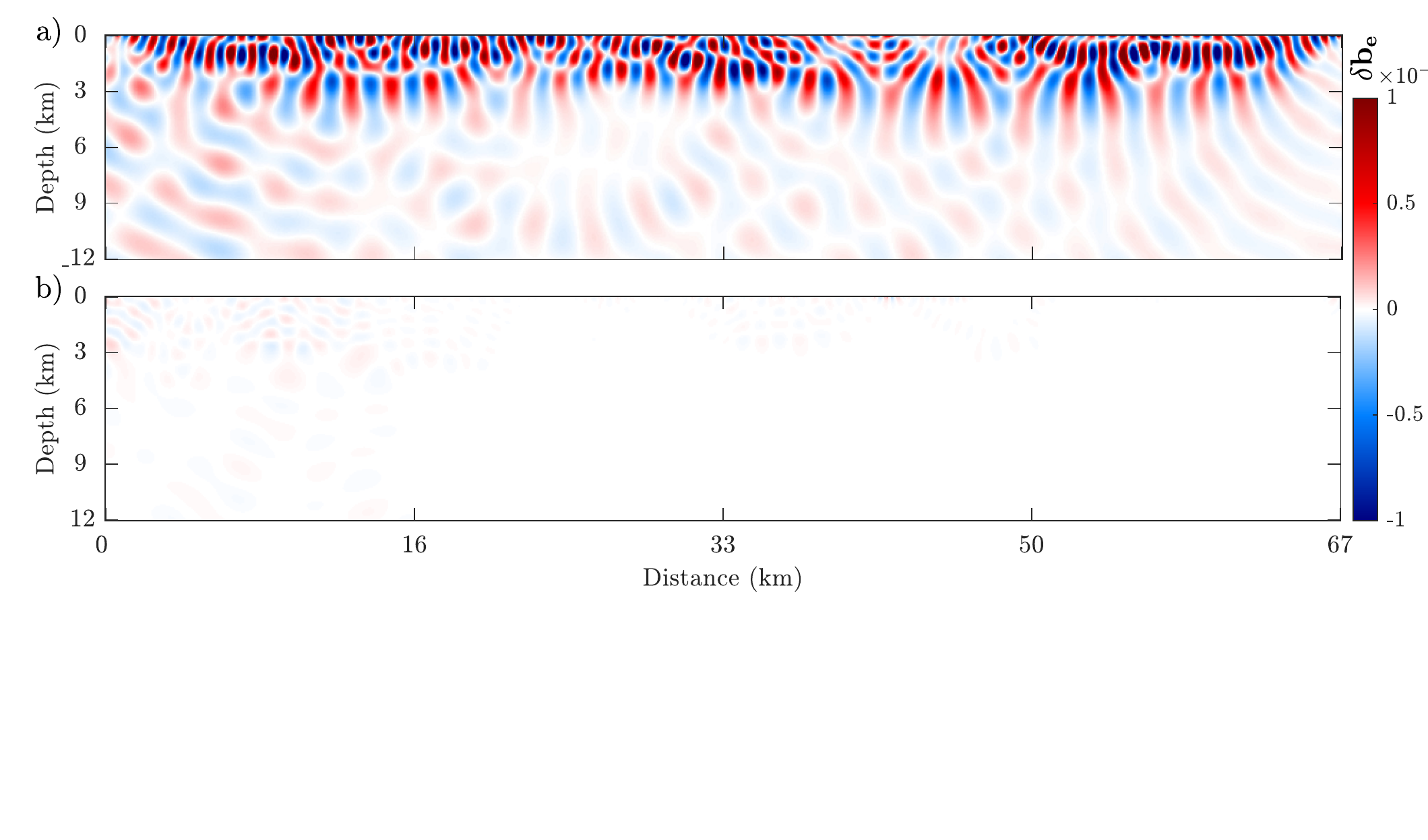} 
\caption{2004 BP salt model benchmark. (a-b) Estimated scattering source during the first frequency batch inversion of the first path (a) iteration 1. (b) iteration 10. The physical source is located at (x,z)=(7500,1350) m.}
\label{fig_deltab}
\end{center}
\end{figure}

\begin{figure}[htb!]
\centering
\begin{center}
\includegraphics[width=17cm,clip=true,trim=0cm 0cm 0cm 0cm]{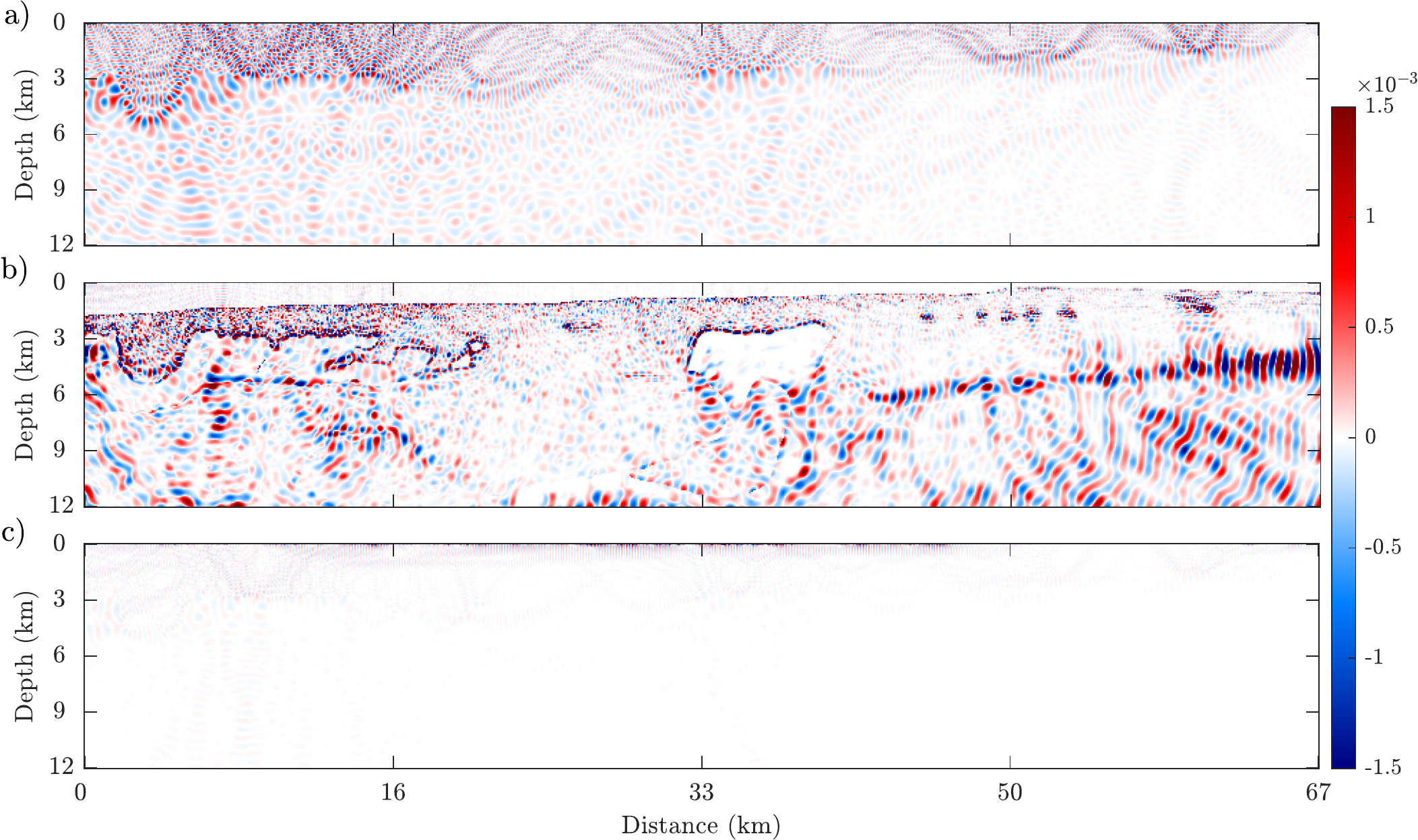} 
\caption{2004 BP salt model benchmark. (a,c) Estimated scattering source during the last frequency batch inversion (5~Hz) of the first path. (a) Iteration 1. (c) Iteration 10. (b) Exact scattering source, $\delta \bold{b}(\bold{m}^*,\bold{m}_{0,5})$ where $\bold{m}_{0,5}$ denotes the starting model of the frequency batch inversion involving the 5~Hz frequency. The physical source is located at (x,z)=(7500,1350) m.}
\label{fig_deltab5}
\end{center}
\end{figure}

The dimensions of the model are 67~km $\times$ 12~km. The ultra long-offset stationary-recording acquisition consists of 67 hydrophones spaced 1 km apart on the seabed, and 450 pressure sources spaced 150 m apart at 25-m depth. For the sake of computational efficiency, we use the spatial reciprocity of Green's functions to process sources as receivers and vice versa. A free-surface boundary condition is used on top of the grid, hence the surface multiples are involved in ES-FWI, which improves illumination while increasing nonlinearity due to more complex data anatomy (Figure~\ref{fig_sismos}). The source signature is a 4~Hz Ricker wavelet. 
The starting model $\bold{m}_0$ is a crude velocity-gradient model (Figure~\ref{fig_bpsalt}b). Direct comparison between the simulated seismograms in the ground-truth model $\bold{m}^*$ and $\bold{m}_0$, namely $\bold{d}^*=\bold{S}(\bold{m}^*) \bold{b}$ and $\bold{d}(\bold{m}_0)=\bold{S}(\bold{m}_0) \bold{b}$, respectively, shows obvious cycle skipping (Figure~\ref{fig_sismos}a), while 
direct comparison between $\bold{d}^*$ and the simulated data-assimilated seismograms in $\bold{m}_0$, namely $\bold{d}^e(\bold{m}_0)=\bold{S}(\bold{m}_0) \left(\bold{b}+\delta \bold{b}^e(\bold{m}_0)\right)$, highlights the ability to match arbitrarily well the recorded data provided that the normal system for $\delta \bold{d}^e$, equation~\ref{eqnormal}, is solved accurately (Figure~\ref{fig_sismos}b). Note that we compute $\bold{d}^e$ in the time domain (Figure~\ref{fig_sismos}b) by solving the normal system "exactly" in the frequency domain with a direct solver for each frequency sampling the source bandwidth before taking the inverse Fourier transform of the monochromatic data. \\
\noindent We perform ES-FWI in the frequency domain in the 1.5 Hz - 5 Hz frequency band with a frequency interval of 0.25~Hz. Following a multiscale frequency continuation strategy, we proceed over small frequency batches from low frequencies to higher ones when each batch contains two frequencies with one frequency overlap. We perform three paths through the frequencies, using the final model of one path as the initial model of the next one. The starting and finishing frequencies of the three paths are [1.5, 2], [1.5, 4], [3, 5] Hz, respectively. The stopping criterion of iterations is a maximum of ten iterations per batch in all cases. We stabilize inversion with bound constraints and compound regularization formulated as the infimal convolution of TV and Tikhonov regularizations \citep{Aghamiry_2019_CRO}. \\
\noindent The final ES-FWI model denoted by $\bold{m}_{f}^{e}$ matches fairly well the ground-truth model with some mild ringing artifacts (Figure~\ref{fig_bpsalt}a,c). Accordingly, the seismograms computed in this final model $\bold{d}(\bold{m}_f^{e})= \bold{S}(\bold{m}_f^{e}) \bold{b}$ match fairly well the recorded counterparts $\bold{d}^*$ (Figure~\ref{fig_sismos}c). Since ES-FWI aims at fitting both the data and the source, it is also worth checking how the wave equation constraint is relaxed at the beginning of the inversion and how well it is satisfied at the convergence point. Example of scattering sources at the first and final iterations of the first frequency-batch inversion, namely $\sum_s \delta \bold{b}_s^e(\bold{m}_{0,1.5}^e)$ and $\delta \bold{b}_s^e(\bold{m}_{10,1.5}^e)$ respectively, are shown in Figure~\ref{fig_deltab} where the second subscript denotes the frequency. Matching the data with the highly-inaccurate model shown in Figure~\ref{fig_bpsalt}b generates strong scattering sources with large spatial support covering the full subsurface domain (Figure~\ref{fig_deltab}a). At iteration 10, we show how these scattering sources vanish as the multivariate inversion for wavefield and parameters approaches the convergence point, equations~\ref{eqm1}-\ref{eqm2}. The fact the scattering sources vanish means that both the physical source and the data were matched since the scattering sources are computed by back-propagating the data residuals.
Another example is shown in Figure~\ref{fig_deltab5} for the 5~Hz frequency and the first frequency path. We show also the true scattering source $\delta \bold{b}(\bold{m}^*,\bold{m}_{0,5})$ in Figure~\ref{fig_deltab5}b for sake of comparison with $\delta \bold{b}_s^e(\bold{m}_{0,5})$ even if it cannot be matched since $\bold{m}^*$ is beyond the resolution power of FWI at 5~Hz. The fact that $\delta \bold{b}(\bold{m}^*,\bold{m}_{0,5})$ mainly contains the edges of the salt bodies suggests that their long wavelengths have been already reconstructed at previous frequencies. Also, the amplitudes of $\delta \bold{b}_s^e(\bold{m}_{0,5})$ (Figure~\ref{fig_deltab5}a) look smaller than those of the 1.5~-Hz scattering source (Figure~\ref{fig_deltab}a) suggesting that the weak scattering approximation may be valid at this stage. Therefore, we may switch to FWI at this stage or use a low-rank approximation of the data-domain Hessian in ES-FWI. These numerical strategies will be discussed in more details in future studies. Finally, we see that $\delta \bold{b}_s^e(\bold{m}_{10,5})$ vanishes at the convergence point (Figure~\ref{fig_deltab5}c).

\section{Conclusion}
ES-FWI modifies the gradient of FWI with more accurate wavefields in the virtual sources of the partial derivative data.  Beyond the Born approximation, this allows the partial derivative data to account approximately for scattering generated by the sought model perturbation along the incident wavepath to the scatterers. These more accurate wavefields are formulated as the sum of the background wavefields plus an approximation of the scattered wavefields by the sought model perturbation. These scattered wavefields are computed by solving the wave equation in the background model with a scattering source in the right-hand side (namely, by solving the Lippmann-Schwinger equation). These scattering sources are estimated by solving an underdetermined scattered data fitting problem, where the scattered data are the data residuals at the current iteration. Accordingly, the scattering sources are computed by propagating backward in time the weighted data residuals by the inverse of the data-domain Hessian. 
When the wavefields and the model are updated in alternating mode, the latter is updated by minimizing the scattering sources in a least-squares sense, which amounts to minimize the data residuals while pushing the updated model toward the ground truth model.
In the end, the imaging principle of FWI and ES-FWI are similar: they rely on the correlation between the simulated virtual sources and the estimated scattering sources from the observables (generally, referred to as the adjoint wavefields in the FWI framework). However, these two ingredients have different expressions in FWI and ES-FWI. The virtual sources are built with data-assimilated wavefields and background wavefields in ES-FWI and FWI, respectively. Moreover, the estimated scattering sources are the least-squares solutions of the scattered-data fitting problem in ES-FWI, while they are the adjoint approximation of this problem in FWI, a consequence of the weak-scattering approximation underlying FWI. These modifications allow ES-FWI to match the recorded data arbitrarily well hence avoiding cycle skipping. However, the recorded scattered data (i.e., the data residuals) cannot be matched accurately by the simulated counterpart (i.e., the partial derivative data) for crude initial models due to the limited resolution with which the scattering sources are estimated. In this case, the data-domain Hessian accounts for the limited resolution with which the simulated scattered data can match the recorded counterpart by reshaping the data residuals in the so-called adjoint source to give more weight to the recorded data at the expense of the simulated data in the weighted data residuals. That is, the descent direction of ES-FWI is mostly driven by the recorded data at the expense of the simulated ones for crude background models. Then,  the full differential information contained in the data residuals is captured more accurately as the inversion converges toward the minimizer, the accuracy of the scattering source estimation improves,  and the data-domain Hessian has a similar effect on the recorded and simulated data, accordingly.
With these remarks, the conditions that are required to converge toward accurate models remain unclear. The accuracy of the reconstructed wavefields degrades away from the receivers as they are built by propagating backward in time the recorded data in the background model. The spatial support and the amplitudes of the scattering sources increase accordingly and it becomes more challenging to minimize their footprint. Therefore, it is likely that heuristic approaches based on layer stripping combined with regularizations are useful to guide the inversion toward accurate models when the inversion starts from crude initial models. The computational burden of ES-FWI generated by the data-domain Hessian of the scattering-source estimation problem is another issue. However, taking into account for the Hessian accurately is probably not necessary during the late stages of the inversion when the weak scattering approximation becomes valid. Therefore, designing numerical strategies providing low-rank approximation of the data-domain Hessian and tuning the accuracy with which this Hessian should be approximated in iterations is the aim of ongoing work. The resilience of the method to noise in the frame of real data applications still needs to be assessed.


\section{Acknowledgments}
This study has been performed in the frame of the WIND project supported by Petrobras, Shell, Total. The authors are grateful to the OPAL infrastructure from Observatoire de la C\^ote d'Azur (CRIMSON) for providing resources and support. This work was granted access to the HPC resources of IDRIS under the allocation 0596 made by GENCI.

\newcommand{\SortNoop}[1]{}


\begin{thebibliography}{68}
\expandafter\ifx\csname natexlab\endcsname\relax\def\natexlab#1{#1}\fi

\bibitem[Abubakar et~al.(2008)Abubakar, Hu, {van den Berg}, \&
  Habashy]{Abubakar_2008_FCS}
Abubakar, A., Hu, W., {van den Berg}, P.~M., \& Habashy, T.~M., 2008.
\newblock A finite-difference contrast source inversion method, {\it Inverse
  problems\/}, {\bf 24}(6), 1--17.

\bibitem[Abubakar et~al.(2009)Abubakar, Hu, Habashy, \& {van den
  Berg}]{Abubakar_2009_FDC}
Abubakar, A., Hu, W., Habashy, T.~M., \& {van den Berg}, P.~M., 2009.
\newblock Application of the finite-difference contrast-source inversion
  algorithm to seismic full-waveform data, {\it Geophysics\/}, {\bf 74}(6),
  WCC47--WCC58.

\bibitem[Abubakar et~al.(2011)Abubakar, Pan, Li, Zhang, Habashy, \& {van den
  Berg}]{Abubakar_2011_TSF}
Abubakar, A., Pan, G., Li, M., Zhang, L., Habashy, T.~M., \& {van den Berg},
  P., 2011.
\newblock Three-dimensional seismic full-waveform inversion using the
  finite-difference contrast source inversion method, {\it Geophysical
  prospecting\/}, {\bf 59}, 874--888.

\bibitem[Aghamiry et~al.(2019{\natexlab{a}})Aghamiry, Gholami, \&
  Operto]{Aghamiry_2019_AMW}
Aghamiry, H., Gholami, A., \& Operto, S., 2019{\natexlab{a}}.
\newblock {ADMM}-based multi-parameter wavefield reconstruction inversion in
  {VTI} acoustic media with {TV} regularization, {\it Geophysical Journal
  International\/}, {\bf 219}(2), 1316--1333.

\bibitem[Aghamiry et~al.(2019{\natexlab{b}})Aghamiry, Gholami, \&
  Operto]{Aghamiry_2019_IBC}
Aghamiry, H., Gholami, A., \& Operto, S., 2019{\natexlab{b}}.
\newblock Implementing bound constraints and total-variation regularization in
  extended full waveform inversion with the alternating direction method of
  multiplier: application to large contrast media, {\it Geophysical Journal
  International\/}, {\bf 218}(2), 855--872.

\bibitem[Aghamiry et~al.(2019{\natexlab{c}})Aghamiry, Gholami, \&
  Operto]{Aghamiry_2019_IWR}
Aghamiry, H., Gholami, A., \& Operto, S., 2019{\natexlab{c}}.
\newblock Improving full-waveform inversion by wavefield reconstruction with
  alternating direction method of multipliers, {\it Geophysics\/}, {\bf 84(1)},
  R139--R162.

\bibitem[Aghamiry et~al.(2020{\natexlab{a}})Aghamiry, Gholami, \&
  Operto]{Aghamiry_2019_AEW}
Aghamiry, H., Gholami, A., \& Operto, S., 2020{\natexlab{a}}.
\newblock Accurate and efficient wavefield reconstruction in the time domain,
  {\it Geophysics\/}, {\bf 85(2)}, A7--A12.

\bibitem[Aghamiry et~al.(2020{\natexlab{b}})Aghamiry, Gholami, \&
  Operto]{Aghamiry_2019_CRO}
Aghamiry, H., Gholami, A., \& Operto, S., 2020{\natexlab{b}}.
\newblock Compound regularization of {full-waveform inversion} for imaging
  piecewise media, {\it {IEEE} Transactions on Geoscience and Remote
  Sensing\/}, {\bf 58}(2), 1192--1204.

\bibitem[Aghamiry et~al.(2020{\natexlab{c}})Aghamiry, Gholami, \&
  Operto]{Aghamiry_2019_VWR}
Aghamiry, H., Gholami, A., \& Operto, S., 2020{\natexlab{c}}.
\newblock Multi-parameter wavefield reconstruction inversion for wavespeed and
  attenuation with bound constraints and total variation regularization, {\it
  Geophysics\/}, {\bf 85(4)}(4), 1--16.

\bibitem[Aghamiry et~al.(2020{\natexlab{d}})Aghamiry, Gholami, \&
  Operto]{Aghamiry_2020_RWI}
Aghamiry, H., Gholami, A., \& Operto, S., 2020{\natexlab{d}}.
\newblock Robust wavefield inversion with phase retrieval, {\it Geophysical
  Journal International\/}, {\bf 221}(2), 1327--1340.

\bibitem[Aghamiry et~al.(2021{\natexlab{a}})Aghamiry, Gholami, \&
  Operto]{Aghamiry_2020_CIT}
Aghamiry, H., Gholami, A., \& Operto, S., 2021{\natexlab{a}}.
\newblock Complex-valued imaging with total variation regularization: An
  application to full-waveform inversion in visco-acoustic, {\it {SIAM Journal
  on Imaging Sciences (SIIMS)}\/}, {\bf 14}(1), 58--91.

\bibitem[Aghamiry et~al.(2021{\natexlab{b}})Aghamiry, Gholami, \&
  Operto]{Aghamiry_2020_FWI}
Aghamiry, H., Gholami, A., \& Operto, S., 2021{\natexlab{b}}.
\newblock {Full Waveform Inversion by Proximal Newton Methods using Adaptive
  Regularization}, {\it Geophysical Journal International\/}, {\bf 224}(1),
  169--180.

\bibitem[Aghamiry et~al.(2021{\natexlab{c}})Aghamiry, Gholami, \&
  Operto]{Aghamiry_2021_OEF}
Aghamiry, H., Gholami, A., \& Operto, S., 2021{\natexlab{c}}.
\newblock On efficient frequency-domain full-waveform inversion with extended
  search space, {\it Geophysics\/}, {\bf 86(2)}(2), R237.

\bibitem[Aghamiry et~al.(2022)Aghamiry, Gholami, \& Operto]{Aghamiry_2022_HWR}
Aghamiry, H., Gholami, A., \& Operto, S., 2022.
\newblock Highly-accurate wavefield reconstruction inversion using convergent
  born series, in {\em Expanded Abstracts\/}, {EAGE}.

\bibitem[Aghamiry et~al.(2021{\natexlab{d}})Aghamiry, Mamfoumbi-Ozoumet,
  Gholami, \& Operto]{Aghamiry_2021_EES}
Aghamiry, H.~S., Mamfoumbi-Ozoumet, F.~W., Gholami, A., \& Operto, S.,
  2021{\natexlab{d}}.
\newblock Efficient extended-search space full-waveform inversion with unknown
  source signatures, {\it Geophysical Journal International\/}, {\bf 227}(1),
  257--274.

\bibitem[Aghazade et~al.(2022{\natexlab{a}})Aghazade, Gholami, Aghamiry, \&
  Operto]{Aghazade_2021_RSS}
Aghazade, K., Gholami, A., Aghamiry, H., \& Operto, S., 2022{\natexlab{a}}.
\newblock Randomized source sketching for full waveform inversion, {\it {IEEE}
  {T}ransactions on {G}eoscience and {R}emote {S}ensing\/}, {\bf 60}, 1--12.

\bibitem[Aghazade et~al.(2022{\natexlab{b}})Aghazade, Gholami, Aghamiry, \&
  Operto]{Aghazade_2022_AAA}
Aghazade, K., Gholami, A., Aghamiry, H., \& Operto, S., 2022{\natexlab{b}}.
\newblock Anderson accelerated augmented lagrangian for extended waveform
  inversion, {\it Geophysics\/}, {\bf 87}, R79--R91.

\bibitem[Baek et~al.(2014)Baek, Calandra, \& Demanet]{Baek_2014_RLS}
Baek, H., Calandra, H., \& Demanet, L., 2014.
\newblock Velocity estimation via registration-guided least-squares inversion,
  {\it Geophysics\/}, {\bf 79}(2), R79--R89.

\bibitem[Baeten et~al.(2013)Baeten, {de Maag}, Plessix, Klaassen, Qureshi,
  Kleemeyer, {ten Kroode}, \& Rujie]{Baeten_2013_ULF}
Baeten, G., {de Maag}, J.~W., Plessix, R.-E., Klaassen, R., Qureshi, T.,
  Kleemeyer, M., {ten Kroode}, F., \& Rujie, Z., 2013.
\newblock The use of low frequencies in a full-waveform inversion and impedance
  inversion land seismic case study, {\it Geophysical Prospecting\/}, {\bf
  61}(4), 701--711.

\bibitem[Banerjee et~al.(2013)Banerjee, Walsh, Aquino, \&
  Bonnet]{Banerjee_2013_LSP}
Banerjee, B., Walsh, T.~F., Aquino, W., \& Bonnet, M., 2013.
\newblock Large scale parameter estimation problems in frequency-domain
  elastodynamics using an error in constitutive equation functional, {\it
  Computational Methods in Applied Mechanics Engineering\/}, {\bf 253}, 60--72.

\bibitem[Barnier et~al.(2012)Barnier, Biondi, Clapp, \&
  Biondi]{Barnier_2022_FWI}
Barnier, G., Biondi, E., Clapp, R.~G., \& Biondi, B., 2012.
\newblock Full waveform inversion by model extension: theory, design and
  optimization, {\it arXiv.2205.14341v1\/}.

\bibitem[Biondi \& Almomin(2014)]{Biondi_2014_SIF}
Biondi, B. \& Almomin, A., 2014.
\newblock Simultaneous inversion of full data bandwidth by tomographic
  full-waveform inversion, {\it Geophysics\/}, {\bf 79(3)}, WA129--WA140.

\bibitem[Brenders et~al.(2022)Brenders, Dellinger, Ahmed, Díaz, Gherasim, Jin,
  Vyas, \& Naranjo]{Brenders_2022_WEL}
Brenders, A., Dellinger, J., Ahmed, I., Díaz, E., Gherasim, M., Jin, H., Vyas,
  M., \& Naranjo, J., 2022.
\newblock The wolfspar experience with low-frequency seismic source field data:
  Motivation, processing, and implications, {\it The Leading Edge\/}, {\bf
  41}(1), 9--18.

\bibitem[Chavent \& Sabatier(1996)]{Chavent_1996_IPW}
eds Chavent, G. \& Sabatier, P.~C., 1996.
\newblock {\it Inverse problems of wave propagation and diffraction\/}, Lecture
  Notes in Physics, Proceedings, Aix-les-Bains, France, Springer.

\bibitem[Engquist et~al.(2016)Engquist, Froese, \& Yang.]{Engquist_2016_WAS}
Engquist, B., Froese, B.~D., \& Yang., Y., 2016.
\newblock Optimal transport for seismic full waveform inversion, {\it
  Communications in Mathematical Sciences\/}, {\bf 14}(8), 2309--2330.

\bibitem[Epanomeritakis et~al.(2008)Epanomeritakis, Akcelik, Ghattas, \&
  Bielak]{Epanomeritakis_2008_NCG}
Epanomeritakis, I., Akcelik, V., Ghattas, O., \& Bielak, J., 2008.
\newblock A {N}ewton-{CG} method for large-scale three-dimensional elastic full
  waveform seismic inversion, {\it Inverse Problems\/}, {\bf 24}, 1--26.

\bibitem[Esser et~al.(2018)Esser, Guasch, van Leeuwen, Aravkin, \&
  Herrmann]{Esser_2018_TVR}
Esser, E., Guasch, L., van Leeuwen, T., Aravkin, A.~Y., \& Herrmann, F.~J.,
  2018.
\newblock Total variation regularization strategies in {F}ull-{W}aveform
  {I}nversion, {\it {SIAM} Journal Imaging Sciences\/}, {\bf 11}(1), 376--406.

\bibitem[Fang et~al.(2018)Fang, Wang, \& Herrmann]{Fang_2018_SEF}
Fang, Z., Wang, R., \& Herrmann, F.~J., 2018.
\newblock Source estimation for wavefield-reconstruction inversion, {\it
  Geophysics\/}, {\bf 83}(4), R345--R359.

\bibitem[Fiddy \& Pommet(1996)]{Fiddy_1996_RSS}
Fiddy, M.~A. \& Pommet, D.~A., 1996.
\newblock {\it Recovery of strongly scattering permitivitty distributions from
  limited backscattered data using a nonlinear filteringtechnique\/}, vol. 15B,
  pp. 58--70, Lecture Notes in Physics, Springer.

\bibitem[Fu \& Symes(2017)]{Fu_2017_DPM}
Fu, L. \& Symes, W.~W., 2017.
\newblock A discrepancy-based penalty method for extended waveform inversion,
  {\it Geophysics\/}, {\bf R282-R298}, 78--82.

\bibitem[Gauthier et~al.(1986)Gauthier, Virieux, \&
  Tarantola]{Gauthier_1986_TDN}
Gauthier, O., Virieux, J., \& Tarantola, A., 1986.
\newblock Two-dimensional nonlinear inversion of seismic waveforms: numerical
  results, {\it Geophysics\/}, {\bf 51}(7), 1387--1403.

\bibitem[Gholami et~al.(2022)Gholami, Aghamiry, \& Operto]{Gholami_2022_EFW}
Gholami, A., Aghamiry, H.~S., \& Operto, S., 2022.
\newblock Extended full waveform inversion in the time domain by the augmented
  {L}agrangian method, {\it Geophysics\/}, {\bf 87}(1), R63--R77.

\bibitem[Goldstein \& Osher(2009)]{Goldstein_2009_SBM}
Goldstein, T. \& Osher, S., 2009.
\newblock The split {B}regman method for {L}1-regularized problems, {\it {SIAM}
  Journal on Imaging Sciences\/}, {\bf 2}(2), 323--343.

\bibitem[Golub \& Pereyra(2003)]{Golub_2013_VPM}
Golub, G. \& Pereyra, V., 2003.
\newblock Separable nonlinear least squares: the variable projection method and
  its applications, {\it Inverse problems\/}, {\bf 19}(2), R1.

\bibitem[G{\'{o}}rszczyk et~al.(2017)G{\'{o}}rszczyk, Operto, \&
  Malinowski]{Gorszczyk_2017_TRW}
G{\'{o}}rszczyk, A., Operto, S., \& Malinowski, M., 2017.
\newblock Toward a robust workflow for deep crustal imaging by {FWI} of {OBS}
  data: The eastern nankai trough revisited, {\it Journal of Geophysical
  Research: Solid Earth\/}, {\bf 122}(6), 4601--4630.

\bibitem[Guo et~al.(2022)Guo, Operto, Gholami, \& Aghamiry]{Guo_2022_PID}
Guo, G., Operto, S., Gholami, A., \& Aghamiry, H.~S., 2022.
\newblock A practical implementation of data-space hessian in the time-domain
  formulation of source extended full-waveform inversion, in {\em Second
  International Meeting for Applied Geoscience \& Energy\/}, pp. 757--761,
  Society of Exploration Geophysicists and American Association of
  Petroleum~….

\bibitem[Hajjaj et~al.(2022)Hajjaj, de~Ridder, Livermore, \&
  Ravasi]{Hajjaj_2022_wavefield}
Hajjaj, R.~F., de~Ridder, S.~A., Livermore, P.~W., \& Ravasi, M., 2022.
\newblock Wavefield reconstruction inversion modelling of marchenko focusing
  functions, {\it arXiv preprint arXiv:2210.14570\/}.

\bibitem[Huang et~al.(2018{\natexlab{a}})Huang, Nammour, \&
  Symes]{Huang_2018_SEW}
Huang, G., Nammour, R., \& Symes, W.~W., 2018{\natexlab{a}}.
\newblock Source-independent extended waveform inversion based on space-time
  source extension: Frequency-domain implementation, {\it Geophysics\/}, {\bf
  83}(5), R449--R461.

\bibitem[Huang et~al.(2018{\natexlab{b}})Huang, Nammour, \&
  Symes]{Huang_2018_VSE}
Huang, G., Nammour, R., \& Symes, W.~W., 2018{\natexlab{b}}.
\newblock Volume source-based extended waveform inversion, {\it Geophysics\/},
  {\bf 83}(5), R369--387.

\bibitem[Lee \& Pyun(2020)]{Lee_2020_SFW}
Lee, D. \& Pyun, S., 2020.
\newblock Seismic full-waveform inversion using minimization of virtual
  scattering sources, {\it Geophysics\/}, {\bf 85}, R299--R311.

\bibitem[Lin et~al.(2022)Lin, Liu, Xing, \& Lin]{Lin_2022_TWR}
Lin, Y., Liu, H., Xing, L., \& Lin, H., 2022.
\newblock Time-domain wavefield reconstruction inversion solutions in the
  weighted full waveform inversion form, {\it IEEE Transactions on Geoscience
  and Remote Sensing\/}, {\bf 60}, 1--14.

\bibitem[Ma \& Hale(2013)]{Ma_2013_WRT}
Ma, Y. \& Hale, D., 2013.
\newblock Wave-equation reflection traveltime inversion with dynamic warping
  and full waveform inversion, {\it Geophysics\/}, {\bf 78}(6), R223--R233.

\bibitem[M\'etivier et~al.(2018)M\'etivier, Allain, Brossier, M\'erigot, Oudet,
  \& Virieux]{Metivier_2018_OTM}
M\'etivier, L., Allain, A., Brossier, R., M\'erigot, Q., Oudet, E., \& Virieux,
  J., 2018.
\newblock Optimal transport for mitigating cycle skipping in full waveform
  inversion: a graph space transform approach, {\it Geophysics\/}, {\bf 83}(5),
  R515--R540.

\bibitem[Mora(1987)]{Mora_1987_NTD}
Mora, P.~R., 1987.
\newblock Nonlinear two-dimensional elastic inversion of multi-offset seismic
  data, {\it Geophysics\/}, {\bf 52}, 1211--1228.

\bibitem[Mulder \& Plessix(2008)]{Mulder_2008_ESI}
Mulder, W. \& Plessix, R.~E., 2008.
\newblock Exploring some issues in acoustic full waveform inversion, {\it
  Geophysical Prospecting\/}, {\bf 56}(6), 827--841.

\bibitem[Nocedal \& Wright(2006)]{Nocedal_2006_NO}
Nocedal, J. \& Wright, S.~J., 2006.
\newblock {\it Numerical Optimization\/}, Springer, 2nd edn.

\bibitem[Parikh \& Boyd(2013)]{Parikh_2013_PA}
Parikh, N. \& Boyd, S., 2013.
\newblock Proximal algorithms, {\it Foundations and Trends in Optimization\/},
  {\bf 1(3)}, 123--231.

\bibitem[Peters \& Herrmann(2019)]{Peters_2019_ANS}
Peters, B. \& Herrmann, F.~J., 2019.
\newblock A numerical solver for least-squares sub-problems in 3d wavefield
  reconstruction inversion and related problem formulations, in {\em SEG
  International Exposition and Annual Meeting\/}, OnePetro.

\bibitem[Pratt et~al.(1998)Pratt, Shin, \& Hicks]{Pratt_1998_GNF}
Pratt, R.~G., Shin, C., \& Hicks, G.~J., 1998.
\newblock {G}auss-{N}ewton and full {N}ewton methods in frequency-space seismic
  waveform inversion, {\it Geophysical Journal International\/}, {\bf 133},
  341--362.

\bibitem[Prunty \& Snieder(2020)]{Prunty_2020_ALI}
Prunty, A.~C. \& Snieder, R.~K., 2020.
\newblock An acoustic lippmann-schwinger inversion method: applications and
  comparison with the linear sampling method, {\it Journal of Physics
  Communications\/}, {\bf 4}(015007), 1--14.

\bibitem[Rizzuti et~al.(2021)Rizzuti, Louboutin, Wang, \&
  Herrmann]{Rizzuti_2021_DFW}
Rizzuti, G., Louboutin, M., Wang, R., \& Herrmann, F.~J., 2021.
\newblock A dual formulation of wavefield reconstruction inversion for
  large-scale seismic inversion, {\it Geophysics\/}, {\bf 86}(6), R879--R893.

\bibitem[Shin et~al.(2001)Shin, Jang, \& Min]{Shin_2001_IAP}
Shin, C., Jang, S., \& Min, D.~J., 2001.
\newblock Improved amplitude preservation for prestack depth migration by
  inverse scattering theory, {\it Geophysical Prospecting\/}, {\bf 49},
  592--606.

\bibitem[Sirgue et~al.(2010)Sirgue, Barkved, Dellinger, Etgen, Albertin, \&
  Kommedal]{Sirgue_2010_FWI}
Sirgue, L., Barkved, O.~I., Dellinger, J., Etgen, J., Albertin, U., \&
  Kommedal, J.~H., 2010.
\newblock Full waveform inversion: the next leap forward in imaging at
  {V}alhall, {\it First Break\/}, {\bf 28}, 65--70.

\bibitem[Symes(2008)]{Symes_2008_MVA}
Symes, W.~W., 2008.
\newblock Migration velocity analysis and waveform inversion, {\it Geophysical
  Prospecting\/}, {\bf 56}, 765--790.

\bibitem[Symes(2020)]{Symes_2020_WRI}
Symes, W.~W., 2020.
\newblock Wavefield reconstruction inversion: an example, {\it Inverse
  Problems\/}, {\bf 36}(10), 105010.

\bibitem[Tarantola(1984)]{Tarantola_1984_ISR}
Tarantola, A., 1984.
\newblock Inversion of seismic reflection data in the acoustic approximation,
  {\it Geophysics\/}, {\bf 49}(8), 1259--1266.

\bibitem[Tarantola \& Valette(1982)]{Tarantola_1982_GNI}
Tarantola, A. \& Valette, B., 1982.
\newblock Generalized nonlinear inverse problems solved using the least square
  criterion, {\it Reviews of Geophysical and Space Physics\/}, {\bf 20},
  219--232.

\bibitem[{van den Berg} \& Kleinman(1997)]{vandenBerg_1997_CSI}
{van den Berg}, P.~M. \& Kleinman, R.~E., 1997.
\newblock A contrast source inversion method, {\it Inverse Problems\/}, {\bf
  13}(6), 1607.

\bibitem[van Leeuwen(2019)]{vanLeeuwen_2019_ANO}
van Leeuwen, T., 2019.
\newblock A note on extended full waveform inversion, {\it arXiv preprint
  arXiv:1904.00363\/}.

\bibitem[{van Leeuwen} \& Herrmann(2016)]{vanLeeuwen_2016_PMP}
{van Leeuwen}, T. \& Herrmann, F., 2016.
\newblock A penalty method for {PDE}-constrained optimization in inverse
  problems, {\it Inverse Problems\/}, {\bf 32(1)}, 1--26.

\bibitem[{van Leeuwen} \& Herrmann(2013)]{VanLeeuwen_2013_MLM}
{van Leeuwen}, T. \& Herrmann, F.~J., 2013.
\newblock Mitigating local minima in full-waveform inversion by expanding the
  search space, {\it Geophysical Journal International\/}, {\bf 195(1)},
  661--667.

\bibitem[{van Leeuwen} et~al.(2014){van Leeuwen}, Herrmann, \&
  Peters]{vanLeeuwen_2014_NTF}
{van Leeuwen}, T., Herrmann, F., \& Peters, B., 2014.
\newblock A new take on {FWI} - wavefield reconstruction inversion, in {\em
  76$^{th}$ Annual EAGE Meeting (Amsterdam\/}.

\bibitem[Virieux \& Operto(2009)]{Virieux_2009_OFW}
Virieux, J. \& Operto, S., 2009.
\newblock An overview of full waveform inversion in exploration geophysics,
  {\it Geophysics\/}, {\bf 74}(6), WCC1--WCC26.

\bibitem[Wang et~al.(2016)Wang, Yingst, Farmer, \& Leveille]{Wang_2016_FIR}
Wang, C., Yingst, D., Farmer, P., \& Leveille, J., 2016.
\newblock Full-waveform inversion with the reconstructed wavefield method, in
  {\em SEG Technical Program Expanded Abstracts\/}, pp. 1237--1241.

\bibitem[Wang et~al.(2017)Wang, Yingst, Farmer, Jones, Martin, \&
  Leveille]{Wang_2017_RFI}
Wang, C., Yingst, D., Farmer, P., Jones, I., Martin, G., \& Leveille, J., 2017.
\newblock Reconstructed full-waveform inversion with the extended source, in
  {\em SEG Technical Program Expanded Abstracts\/}, pp. 1449--1453.

\bibitem[Warner \& Guasch(2016)]{Warner_2016_AWI}
Warner, M. \& Guasch, L., 2016.
\newblock Adaptive waveform inversion: Theory, {\it Geophysics\/}, {\bf 81}(6),
  R429--R445.

\bibitem[Yang et~al.(2018)Yang, Engquist, Sun, \& Hamfeldt]{Yang_2017_AOT}
Yang, Y., Engquist, B., Sun, J., \& Hamfeldt, B.~F., 2018.
\newblock Application of optimal transport and the quadratic {W}asserstein
  metric to full-waveform inversion, {\it Geophysics\/}, {\bf 83}(1), R43--R62.

\bibitem[Yao et~al.(2019)Yao, da~Silva, Warner, Wu, \& Yang]{Yao_2019_TCS}
Yao, G., da~Silva, N.~V., Warner, M., Wu, D., \& Yang, C., 2019.
\newblock Tackling cycle skipping in full-waveform inversion with intermediate
  data, {\it Geophysics\/}, {\bf 84}(3), R411--R427.

\end{thebibliography}
\end{document}